\documentclass[12pt]{article}

\usepackage{amsfonts}
\usepackage{amssymb}
\usepackage{amsmath}
\usepackage{amsthm}
\usepackage{caption}
\usepackage{enumitem}
\usepackage[pdftex]{color,graphicx}
\usepackage{tikz,pgfplots}
\usepackage[normalem]{ulem}
\usepackage{tkz-tab,caption,subcaption,latexsym,lipsum}
\usetikzlibrary{arrows,shapes,trees,backgrounds,patterns,decorations.pathreplacing}
\usepackage[T1]{fontenc}
\usepackage[utf8]{inputenc}
\usepackage{authblk}
\usepackage{appendix}
\usepackage{slashbox}
\usepackage{multicol}
\allowdisplaybreaks

\makeatletter
\newcommand{\inlineitem}[1][]{\ifnum\enit@type=\tw@
    {\descriptionlabel{#1}}
  \hspace{\labelsep}\else
  \ifnum\enit@type=\z@
       \refstepcounter{\@listctr}\fi
    \quad\@itemlabel\hspace{\labelsep}\fi}
\makeatother

\usetikzlibrary{arrows,shapes,trees,backgrounds,patterns,decorations.pathreplacing}

\RequirePackage{natbib}

\theoremstyle{theorem}

\newtheorem{proposition}{Proposition}
\newtheorem{corollary}{Corollary}

\newtheorem{exam}{Example}

\newtheorem{assum}{Assumption}

\newtheorem{rema}{Remark}
\newenvironment{remark}{\begin{rema} \rm }{\hfill $\triangleleft$ \end{rema}}
\newtheorem{defin}{Definition}
\newenvironment{definition}{\begin{defin} \rm }{\end{defin}}

\makeatletter

\renewcommand*{\@seccntformat}[1]{%
  \csname the#1\endcsname.\quad}
\makeatother

\begin{document}
\title{On Decay Centrality\footnote{Part of this project was carried out while the author was at SUTD--MIT International Design Center at Singapore University of Technology and Design supported by grant IDG31300110.}}
\author{\textsc{Nikolas Tsakas}\footnote{University of Cyprus, Department of Economics, PO Box 20537, 1678 Nicosia, Cyprus, tsakas.nikolaos@ucy.ac.cy}\\
\small{\textit{Department of Economics, University of Cyprus}}}
\maketitle

\begin{abstract}
We establish a relationship between decay centrality and two widely used and computationally cheaper measures of centrality, namely degree and closeness. We show that for low values of the decay parameter the nodes with maximum decay centrality also have maximum degree, whereas for high values of the decay parameter they also maximize closeness. For intermediate values, we provide sufficient conditions that allow the comparison of decay centrality of different nodes and we show via numerical simulations that in the vast majority of networks, the nodes with maximum decay centrality are characterized by a threshold on the decay parameter below which they belong to the set of nodes with maximum degree and above which they belong to the set of nodes with maximum closeness. We also propose a simple rule of thumb that ensures a nearly optimal choice with very high probability.
\end{abstract}


\section{Introduction}

The identification of individuals with influential positions in a network is of outmost importance for a number of problems in economics and beyond, as for instance regarding the diffusion of epidemics \citep[][]{PastorSatorrasVespignani2002}, the stability of systems of interconnected banks \citep[][]{Gofman2015} and the development of criminal networks \citep[][]{Liuetal2015}. Different contexts often lead to different measures of \emph{centrality} that capture the level of the said influence more appropriately. Degree centrality is found to be important in problems of adoption with word--of--mouth communication \citep[][]{GaleottiGoyal2009} and biases in the perception of social norms \citep[][]{Jackson2016}. Katz--Bonacich centrality \citep[][]{Katz1953,Bonacich1987} is found to be crucial in problems related to criminal behavior \citep[][]{Ballesteretal2006}, whereas eigenvector centrality is found to be important in diffusion processes \citep[][]{Banerjeeetal2013}. 

In this article we focus on \emph{Decay Centrality}. This is a measure of centrality in which a node is rewarded for how close it is to other nodes, but in a way that very distant nodes are weighted less than closer ones \citep[see][]{Jackson2008}. It is defined as $\sum\limits_{j\neq i}\delta^{d(i,j)}$, where $0<\delta<1$ is a decay parameter and $d(i,j)$ is the geodesic distance between nodes $i$ and $j$. For low values of $\delta$ decay centrality puts much more weight on closer nodes, thus becoming proportional to degree centrality, whereas for high values of $\delta$ it measures the size of the component a node lies in

It is considered to be richer than other distance related measures, because it captures the idea that the importance of a node for another is proportional to their distance \citep[see for instance][]{JacksonWolinsky1996}. It has been considered important in problems of optimal targeting selection in networks \citep[see][]{Banerjeeetal2013,ChatterjeeDutta2015,Tsakas2016}.  In particular, \cite{ChatterjeeDutta2015} and \cite{Tsakas2016} find decay centrality to be the measure that helps selecting the node that can lead to the maximum diffusion of a given action in a social network.

Nevertheless, its use is cumbersome for two main reasons. First, except in very simple structures, the nodes with maximum decay centrality cannot be easily identified, since the measure depends vastly on the exact network topology and the value of the decay parameter. Second, calculating the decay centrality of all nodes and subsequently choosing the one that maximizes it might be computationally costly, since it requires calculating the geodesic distance between each pair of nodes and subsequently summing a function of them.\footnote{For a network with $n$ nodes, the time complexity for calculating degree and closeness centrality are in $O(n^2)$ and $O(n^3)$ respectively \citep[see][]{BrandesErlebach2005}, where for the calculation of shortest paths that is necessary for closeness centrality is used the simple Dijkstra algorithm \citep[see][]{Dijkstra1959}. Once the shortest paths have been calculated, decay centrality requires the calculation of $\delta^{d(i,j)}$ for each pair of nodes $(i,j)$. Hence, the time complexity of calculating decay centrality is in $O(n^5)$.}

The aim of this article is to show the close connection between decay centrality and two well--studied and computationally cheaper measures, namely degree and closeness centrality. The relations are established both analytically and numerically and suggest that the nodes with maximum decay centrality usually belong either to the set of nodes with maximum degree or to the set of nodes with maximum closeness.

In particular, focusing on connected networks, we show that for sufficiently low values of the decay parameter the nodes that maximize decay centrality belong to the set of nodes with maximum degree, whereas for sufficiently high values of the decay parameter the nodes that maximize decay centrality maximize closeness as well. The first proposition is not surprising as it is already known that for low values of $\delta$ decay centrality is proportional to degree. However, the second proposition establishes a novel relationship between decay and closeness centrality for high values of $\delta$, for which so far decay centrality was associated only with the size of the component a node lied in.

Furthermore, we provide two conditions that are sufficient to order a pair of nodes with respect to their decay centrality for all values of the decay parameter. The conditions depend, in an intuitive way, on the distances between the nodes under comparison and all other nodes in the network.

Finally, we provide a more general relation between decay centrality, degree and closeness that extends to intermediate values of the decay parameter. In particular, we provide sufficient conditions for a node with higher degree than another node to also have higher decay centrality for all $\delta\leq 1/2$ and similarly for a node with higher closeness to also have higher decay centrality for all $\delta\geq1/2$.

Based on those theoretical findings, we attempt to obtain a better understanding on these relations for intermediate values of the decay parameter via numerical simulations. We find that in the vast majority of cases the nodes with maximum decay centrality belong either to the set of nodes with maximum degree or to the set of nodes with maximum closeness. When the two sets intersect a node that belongs to their intersection almost always maximizes decay centrality as well, whereas when the two sets do not intersect for low values of $\delta$ decay centrality is maximized by nodes with maximum degree and as $\delta$ increases there is a threshold above which decay centrality is maximized by nodes with maximum closeness. The threshold varies with the network parameters, however a rule of thumb with a threshold at $\delta=0.5$ is sufficient to ensure that the chosen node is ranked among the top nodes in terms of decay centrality, with high probability.

The relation between different measures of centrality has attracted research interest over the years \citep[see][and references therein]{Valenteetal2008}. For instance, \cite{Faust1997} found strong correlation among centrality measures that included degree and closeness, but not decay centrality, in a network of relations between CEOs. Similarly, \cite{Rothenbergetal1995} found strong concordance in the ranking of individuals across different network centrality measures in a network of individuals who participated in activities with the risk of HIV transmission. Finally, \cite{Valenteetal2008} find similar results, although not as strong correlation coefficients as the previous studies, for a variety of networks corresponding to different process. All of these studies show particularly strong connection between degree and closeness, but do not consider decay centrality. 

More recently, and closer to the flavor of our analysis, \cite{Blochetal2016} provided an axiomatization of several centrality measures, including decay centrality, and showed that all of them can be described using the same set of axioms. Given these axioms, the measures differ only in that they consider different vectors of data that describe the position of nodes in the network. The authors also provide correlation coefficients between different measures in simulated networks and their results are in line with our findings. The simulations report results on correlation coefficients, similarly to the previously mentioned articles, whereas the emphasis of our analysis is put particularly on nodes that are highly ranked in different measures. To some extent our analysis is related to the problem of identifying networks for which different centrality measures generate the same ranking, which remains largely unexplored for non--tree networks. For degree and closeness centrality \cite{Konigetal2014} show that this is indeed the case for nested--split graphs. Finally, there is a broad connection with the literature related to the axiomatization of centrality measures, although not decay centrality per se, and other related problems \citep[see][]{DequiedtZenou2014,PalaciosHuertaVolij2004}.

\section{Notation and Analytical Results}

Consider a set of nodes $N$, with cardinality $n$, which are connected through a network. A network is represented by a family of sets $\mathcal{N}:=\{N_i\subseteq N\ |\ i=1,\dots,n\}$, with $N_i$ denoting the set of nodes that are directly connected with $i$. $N_i$ is called $i$'s neighborhood and has cardinality, $|N_i|$. We focus on undirected networks, where $j \in N_i$ if and only if $i \in N_j$. A path in a network between nodes $i$ and $j$ is a sequence $i_1,...,i_K$ such that $i_1=i$, $i_K=j$ and $i_{k+1} \in N_{i_k}$ for $k=1,...,K-1$. The \emph{geodesic distance}, $d(i,j)$, between two nodes is the length of the shortest path between them. Two nodes are connected if there exists a path between them. The network is connected if every pair of nodes is connected. Our analysis focuses on connected networks. 

For a given network $\mathcal{N}$, a \emph{centrality measure} is a function that maps from the set of nodes to the real numbers, i.e. ${\bf c}: N \to \mathbb{R}$, where $c_i$ is the centrality of node $i$ in the network $\mathcal{N}$. 

The \emph{degree centrality} (or simply \emph{degree}) maps from each node to the cardinality of its neighborhood, i.e. $D_i=|N_i|$. We also define the set of nodes with maximum degree, i.e. $I_{deg}=\arg \max\limits_{i \in N} D_i$. Similarly to this, denote by $D_i^l$ the number of nodes that have geodesic distance $l$ from node $i$, i.e. $D_i^l=|N_i^l|$ where $N_i^l=\{j \in N\ |\ d(i,j)=l\}$. The superscript $l$ will often be omitted when referring to $l=1$ (degree). Also, for connected networks it holds by definition that $\sum\limits_{l=1}^{n-1}N_i^l=n-1$.

The {\em closeness centrality} (or simply {\em closeness}) maps from each node to the inverse of the sum of the geodesic distances from each other agent in the network, i.e. $C_i=\frac{1}{\sum\limits_{j \neq i} d(i,j)}$. Notice that closeness centrality measures how easily a node can reach all other nodes in the network. According to this definition, we define the set of nodes with maximum closeness centrality, i.e. $I_{clos}=\arg \max\limits_{i \in N} C_i$. 

For some of the analytical results we focus on the inverse of closeness, which is known as \emph{farness} and is defined as $F_i=\sum\limits_{j \neq i} d(i,j)$. This is essentially a measure of discentrality, as it measures how far a node is from all other nodes. Obviously the order of nodes with respect to closeness is the reverse of that with respect to farness, hence central nodes are going to be considered those that minimize farness.

Finally, given a decay parameter $\delta \in (0,1)$, \emph{decay centrality} is a measure that maps from each node to the sum of distances from each other node in the network, adjusted by a decay parameter that makes distant nodes count less than closer ones, i.e. $DC_i^{\delta}=\sum\limits_{j\neq i}\delta^{d(i,j)}$. As in the previous two cases, for each value of $\delta$, we define the set of nodes with maximum decay centrality, i.e. $I_{dc}^{\delta}=\arg \max\limits_{i \in N} DC_i^{\delta}$.

\begin{proposition}
\label{proplowdelta}
Exists $\underline{\delta}$ such that for all $\delta \in (0,\underline{\delta})$ and $i,j \in N$, if $D_i>D_j$ then $DC_i(\delta)>DC_j(\delta)$.
\end{proposition}

The proposition states that for low values of the decay parameter a node with higher degree than another will necessarily have higher decay centrality as well. This result implies immediately Corollary \ref{corlowdelta} which states that for sufficiently low values of the decay parameter the set of nodes with maximum of decay centrality will be a subset of the set of nodes with maximum degree.

\begin{corollary}
\label{corlowdelta}
Exists $\underline{\delta}$ such that for all $\delta \in (0,\underline{\delta})$ holds that $I_{dc}^{\delta}\subseteq I_{deg}$.
\end{corollary}

The reason why the two sets may not be identical is that two nodes with the same degree might have different decay centralities. When the degrees of two nodes are equal, differences arise from the factors corresponding to more distant nodes, which previously had a negligible effect. In fact, again for low values of $\delta$, between two nodes with equal degrees, higher decay centrality has the node with higher number of nodes at distance equal to 2, i.e. $D_i^2$. If these quantities are also equal then we move to $D_i^3$ etc. In order to state this result formally we need an additional definition.

\begin{definition}
Given two vectors of real numbers $\mathcal{A}_i=(A_i^1,A_i^2,\dots,A_i^{n-1})$ and $\mathcal{A}_j=(A_j^1,A_j^2,\dots,A_j^{n-1})$ the first vector is larger than the second one according to the \emph{lexicographic order}, i.e. $\mathcal{A}_i >_L \mathcal{A}_j$, if $A_i^l>A_j^l$ for the first $l$ for which $A_i^l$ and $A_j^l$ differ. The two vectors are equal according to the lexicographic order, i.e. $\mathcal{A}_i =_L \mathcal{A}_j$, if $A_i^l=A_j^l$  for all $l \in \{1,\dots,n-1\}$.
\end{definition}

It is apparent that two vectors that are equal according to the lexicographic order, i.e. $D_i^l=D_j^l$ for all $l$, will also have equal decay centralities. The following result covers the remaining cases.

\begin{proposition}
\label{proplowdeltageneral}
Exists $\underline{\delta}$ such that for all $\delta \in (0,\underline{\delta})$ and $i,j \in N$, if $\mathcal{D}_i >_L \mathcal{D}_j$ then $DC_i(\delta)>DC_j(\delta)$, where $\mathcal{D}_i=(D_i,D_i^2,\dots,D_i^{n-1})$
\end{proposition}

Propositions \ref{proplowdelta} and \ref{proplowdeltageneral} allow the comparison of decay centralities between any pair of nodes for low values of $\delta$. We proceed in a similar way to obtain the respective results for high values of $\delta$.

\begin{proposition}
\label{prophighdelta}
Exists $\overline{\delta}$ such that for all $\delta \in (\overline{\delta},1)$ and $i,j \in N$, if $C_i>C_j$ then $DC_i(\delta)>DC_j(\delta)$.
\end{proposition}

Proposition \ref{prophighdelta} establishes a relation between the order of nodes with respect to decay centrality and closeness for high values of the decay parameter. This also implies immediately Corollary \ref{corhighdelta} which states that for sufficiently high values of the decay parameter the set of nodes with maximum decay centrality will be a subset of the set of nodes with maximum closeness.

\begin{corollary}
\label{corhighdelta}
Exists $\overline{\delta}$ such that for all $\delta \in (\overline{\delta},1)$ holds that $I_{dc}^{\delta}\subseteq I_{clos}$.
\end{corollary}

Similarly to the previous case, Proposition \ref{prophighdelta} does not allow the comparison between nodes with equal closeness centralities. In order to do so it is necessary to define for each node the vector $\mathcal{F}_i=(F_i^1,F_i^2,\dots,F_i^{n-1})$, with typical element $F_i^k=(-1)^{k-1}\sum\limits_{l=k}^{n-1}\binom lk D_i^l$, where $0!=1$ by definition. Observe that $F_i^1=\sum\limits_{k \neq i} d(i,k)$, i.e. the \emph{farness} of node $i$. Note that, the rest of the terms of vector $\mathcal{F}$ are related to the derivatives of farness, which is a different interpretation compared to $\mathcal{D}$.\footnote{The alternating signs appear due to the following result: Let $f,g$ continuously differentiable functions in $\mathbb{R}$ such that $f(1)=g(1)$, $f^{(l)}(1)=g^{(l)}(1)$ for all $l \in \{1,\dots,k\}$ and $f^{(k+1)}(1)>g^{(k+1)}(1)$. Then, there is $\overline{x}$ such that for all $x \in (\overline{x},1)$ it holds that $f(x)>g(x)$ if $k$ is even and $f(x)<g(x)$ if $k$ is odd. $f^{(l)}$ denotes $l$-th order derivative of $f$.}

Based on this definition, one can also define the vector $\mathcal{C}_i=(C_i^1,C_i^2,\dots,C_i^{n-1})$, where $C_i^k=1/F_i^k$ if $F_i^k\neq0$ and $C_i^k=0$ if $F_i^k=0$ for all $k \in \{1,\dots,n-1\}$. Note that $C_i^1$ corresponds to the closeness centrality, thus the overscript will often be omitted, and the following equivalence relations hold always.

\begin{remark}
\label{remarkfarclos}
$\mathcal{F}_i >_L \mathcal{F}_j \Leftrightarrow \mathcal{C}_i <_L \mathcal{C}_j$ and $\mathcal{F}_i =_L \mathcal{F}_j \Leftrightarrow \mathcal{C}_i =_L \mathcal{C}_j$.
\end{remark}

It is important to underline that such an equivalence does not hold for all orders. For instance, it does not hold for the \emph{unsorted dominance order} that is defined and used later in the article.

Having said that,  we are ready to provide a result that determines the complete order among nodes in terms of decay centrality.

\begin{proposition}
\label{prophighdeltageneral}
Exists $\overline{\delta}$ such that for all $\delta \in (\overline{\delta},1)$ and $i,j \in N$, if $\mathcal{C}_i >_L \mathcal{C}_j$ then $DC_i(\delta)>DC_j(\delta)$, where $\mathcal{C}_i=(C_i,C_i^2,\dots,C_i^{n-1})$
\end{proposition}

The results so far have established a clear relation between decay centrality, degree and closeness, in the two limits, which provides a more intuitive picture of the characteristics of nodes with high decay centrality for some values of the decay parameter. However, the established relations still leave unanswered what happens for intermediate values of $\delta$.

\begin{definition}
Given two vectors of real numbers $\mathcal{A}_i=(A_i^1,A_i^2,\dots,A_i^{n-1})$ and $\mathcal{A}_j=(A_j^1,A_j^2,\dots,A_j^{n-1})$ the first one is larger than the second one according to the \emph{unsorted dominance order}, i.e. $\mathcal{A}_i >_{UD} \mathcal{A}_j$, if $\sum\limits_{l=1}^k A_i^l\geq\sum\limits_{l=1}^k A_j^l$ for all $k \in \{1,\dots,n-1\}$ and at least one of the inequalities is strict.
\label{defunsorteddominance}
\end{definition}

We use the term \emph{unsorted} because the standard definition of dominance order considers vectors whose elements are already sorted in decreasing order, which is not the case here. Moreover, notice that \emph{unsorted dominance} is a partial order, thus it may not allow the comparison between all nodes. Finally, observe that, under certain assumptions on the vectors under comparison,  the unsorted dominance order can be seen as a deterministic analog of \emph{first order stochastic dominance}. 

Definition \ref{defunsorteddominance} allows us to establish the following result:

\begin{proposition}
For $i,j \in N$, if $\mathcal{D}_i >_{UD} \mathcal{D}_j$ then $DC_i(\delta)>DC_j(\delta)$ for all $\delta \in (0,1)$.
\label{propintermediatedeltadomdeg}
\end{proposition}

Despite not providing a complete order, the result grasps the important relation between degree and closeness in the calculation of decay centrality. On one hand a high degree is beneficial as it is included in all sums $\sum\limits_{l=1}^k D_i^l$, on the other hand, high closeness implies, though indirectly, that a higher number of nodes will be in a shorter distance from $i$, thus boosting all relevant sums upwards.

A similar result can be established using the relation between $\mathcal{F}_i$ and $\mathcal{F}_j$ as follows:

\begin{proposition}
For $i,j \in N$, if $\mathcal{F}_j >_{UD} \mathcal{F}_i$ then $DC_i(\delta)>DC_j(\delta)$ for all $\delta \in (0,1)$.
\label{propintermediatedeltadomfar}
\end{proposition}

Observe that the subscripts in the first relation are reversed compared to Proposition \ref{propintermediatedeltadomdeg}, meaning that lower values of the elements of $\mathcal{F}_i$ are associated with higher decay centrality. Additionally, as it has already been mentioned, there is no equivalence relation between $\mathcal{F}$ and $\mathcal{C}$ with respect to \emph{unsorted dominance}, thus the result needs to be stated using the vector associated with farness, rather than the one associated with closeness.

The next result, builds upon Propositions \ref{proplowdelta} and \ref{proplowdeltageneral}, as it provides several sufficient conditions that characterize nodes with high decay centrality for values of $\delta<1/2$. Namely,

\begin{proposition}
\label{propdeltalowerhalf}
Consider two distinct nodes $i,j \in N$ and let $A_1=D_i-D_j$ and $A_l=D_i^l-D_j^l$ for all $l \in \{2,\dots,n-1\}$. Then, if $A_1>0$ either of the following four conditions is sufficient to ensure that $DC_i(\delta)>DC_j(\delta)$ for all $\delta \in (0,1/2]$.
\begin{multicols}{2}
\begin{enumerate}
\item $2A_1 \geq (n-1)-D_j$
\item $4A_1+2A_2\geq(n-1)-(D_j+D_j^2)$
\item $A_1\geq \max\{|A_2|,\dots,|A_{n-1}|\}$
\item $A_1 \geq \max \left\{\left|\sum\limits_{l=1}^2A_l\right|,\dots,\left|\sum\limits_{l=1}^{n-2}A_l\right|\right\}$
\end{enumerate}
\end{multicols}
\end{proposition}

The essence of all conditions is that as long as node $i$ has sufficiently many more immediate neighbors than node $j$ (high $A_1$) then node $i$ has higher decay centrality not only very close to the limit, but for a large range of values of $\delta$. The first two conditions provide also a simple illustration of the higher relative importance of closer neighbors compared to more distant ones, which is essential for decay centrality. A closer look at the proof of Proposition \ref{propdeltalowerhalf} reveals that the conditions are relatively demanding and the bounds are not tight, which is to be expected given that they hold for any potential network structure. 

The next result establishes similar conditions for high values of $\delta$, where high decay centrality is more closely associated with low farness. Namely,

\begin{proposition}
\label{propdeltaupperhalf}
Consider two distinct nodes $i,j \in N$ and let $B_1=F_i-F_j$ and $B_l=F_i^l-F_j^l$ for all $l \in \{2,\dots,n-1\}$. Then, if $B_1<0$ either of the following two conditions is sufficient to ensure that $DC_i(\delta)>DC_j(\delta)$ for all $\delta \in [1/2,1)$.
\begin{multicols}{2}
\begin{enumerate}
\item $|B_1|\geq \max\{|B_2|,\dots,|B_{n-1}|\}$
\item $|B_1| \geq \max \left\{\left|\sum\limits_{l=1}^2B_l\right|,\dots,\left|\sum\limits_{l=1}^{n-2}B_l\right|\right\}$
\end{enumerate}
\end{multicols}
\end{proposition}

Propositions \ref{propdeltalowerhalf} and \ref{propdeltaupperhalf} provide a connection between decay centrality, degree and closeness for values that extend away from the limits. An important observation is that $\delta=1/2$ is the maximum upper bound for which conditions 3 and 4 of Proposition \ref{propdeltalowerhalf} guarantee the order of decay centralities, given that the remaining conditions hold. Similarly, $\delta=1/2$  is also the minimum lower bound for which conditions 1 and 2 of Proposition \ref{propdeltaupperhalf} guarantee the order of decay centralities, given that the remaining conditions hold. This last observation provides a natural motivation for the rule--of-thumb that is proposed in the next section.

\section{Numerical Results}

The theoretical results have provided several conditions that support the relation between decay centrality, degree and closeness. Nevertheless, some of these conditions may not be easily satisfied in certain networks. For this reason, we perform an extensive set of simulations that intends to identify how strong this relation is in general networks. In particular, we focus on analyzing the extent to which the set of nodes with maximum decay centrality, $I_{dc}^{\delta}$, coincides with either $I_{deg}$ or $I_{clos}$ (or both) for different values of the decay parameter $\delta$.

We simulate undirected \`Erdos-Renyi networks \citep{ErdosRenyi1959}, $G(n,p)$, where $n$ is the network size and $p$ is the link probability. The networks are required to be connected so that geodesic distances are well defined. We consider five network sizes, $n\in\{10,20,50,100,200\}$ and ten link probabilities $p\in\{0.05,0.1,\dots,0.45,0.5\}$ and perform 10000 trials for each combination $(n,p)$.\footnote{The main body of the article contains representative figure and tables. The Online Appendix contains results for $(n,p)=(1000,0.05)$, some additional figures and some extensions. Additional figures are available upon request.}

The first question is how often $I_{dc}^{\delta}$, i.e. the set of nodes with maximum decay centrality, intersects with either $I_{deg}$, i.e. the set of nodes with maximum degree, or with $I_{clos}$, i.e. the set of nodes with maximum closeness. We find that in the vast majority of the cases $I_{dc}^{\delta} \subseteq \left(I_{deg} \cup I_{clos}\right)$ for almost all values of $\delta$ and not only for the limit values. 

It is important to mention that $I_{deg}$ intersects with $I_{clos}$ quite often for random networks. The reasons why this occurs are outside the scope of this paper, however it has an apparent effect on our results as it provides a natural connection between the two limit cases explored by theory. In fact, in ``almost all'' cases where there are nodes that belong both to $I_{deg}$ and $I_{clos}$, those nodes also belong to $I_{dc}^{\delta}$. This result cannot be generalized as there are cases in which, for some values of $\delta$, the nodes with maximum decay centrality do not belong to either $I_{deg}$ or $I_{clos}$. Nevertheless, as it becomes apparent from Tables~\ref{table::coinciding} and~\ref{table::coincidingandintermediate} the frequency with which such cases arise is practically negligible.

\begin{table}[h!]
\centering
 \begin{tabular}{| c || c | c | c | c | c | c | c | c | c | c |} 
 \hline
 \backslashbox{n}{p} & 0.05 & 0.10 & 0.15 & 0.20 & 0.25 & 0.30 & 0.35 & 0.40 & 0.45 & 0.50 \\
 \hline\hline
 10 & 8620 & 8756 & 8951 & 9035 & 9341 & 9570 & 9752 & 9879 & 9971 & 9987\\ 
 \hline
 20 & 6600 & 7229 & 8113 & 8711 & 9349 & 9752 & 9969 & 9998 & 10000 & 10000\\ 
 \hline
 50 &  5921 & 7620 & 8220 & 9330 & 9964 & 10000 & 10000 & 10000 & 10000 & 10000\\ 
 \hline
 100 & 6783 & 7401 & 8794 & 9989 & 10000 & 10000 & 10000 & 10000 & 10000 & 10000\\ 
 \hline
 200 & 7209 & 7593 & 9976 & 10000 & 10000 & 10000 & 10000 & 10000 & 10000 & 10000\\
 \hline
\end{tabular}
\caption{Frequency of occassions where $I_{deg} \cap I_{clos} \neq \emptyset$.}
\label{table::coinciding}
\end{table}

\begin{table}[h!]
\centering
 \begin{tabular}{| c || c | c | c | c | c | c | c | c | c | c |} 
 \hline
 \backslashbox{n}{p} & 0.05 & 0.10 & 0.15 & 0.20 & 0.25 & 0.30 & 0.35 & 0.40 & 0.45 & 0.50 \\
 \hline\hline
 10 & 0 & 0 & 0 & 0 & 0 & 0 & 0 & 0 & 0 & 0\\ 
 \hline
 20 & 11 & 2 & 0 & 0 & 0 & 0 & 0 & 0 & 0 & 0\\ 
 \hline
 50 & 30 & 0 & 0 & 0 & 0 & 0 & 0 & 0 & 0 & 0\\ 
 \hline
 100 & 14 & 0 & 0 & 0 & 0 & 0 & 0 & 0 & 0 & 0\\ 
 \hline
 200 & 0 & 0 & 0 & 0 & 0 & 0 & 0 & 0 & 0 & 0\\
 \hline
\end{tabular}
\caption{Frequency of occassions where $I_{deg} \cap I_{clos} \neq \emptyset$ and $I_{dc}^{\delta} \nsubseteq (I_{deg} \cap I_{clos})$ for some value of $\delta$.}
\label{table::coincidingandintermediate}
\end{table}

\begin{figure}[htbp]
\centering
\makebox[\linewidth][c]{%
\begin{subfigure}[b]{.54\textwidth}
        \centering
        \resizebox{\linewidth}{!}{
\includegraphics{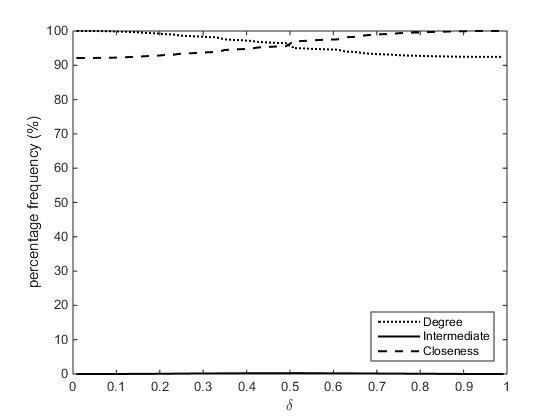}
}
\caption{Including cases where $I_{deg} \cap I_{clos} \neq \emptyset$}
\end{subfigure}
\begin{subfigure}[b]{.54\textwidth}
        \centering
        \resizebox{\linewidth}{!}{
\includegraphics{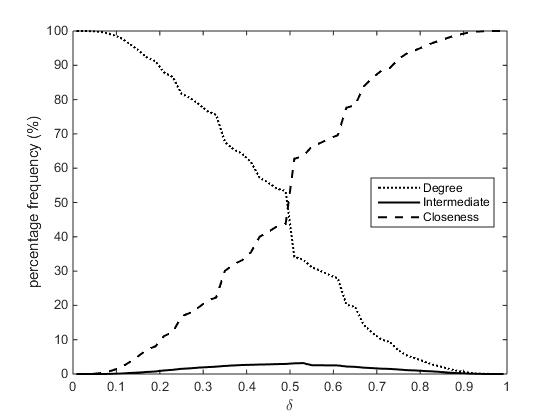}
}
\caption{Excluding cases where $I_{deg} \cap I_{clos} \neq \emptyset$}
\end{subfigure}
}\\
\caption{The dotted (dashed) line shows the frequency with which $I_{dec}^{\delta} \subseteq I_{deg}$ ($I_{dec}^{\delta} \subseteq I_{clos}$), whereas the solid line shows the frequency with which it does not belong to any of the two sets. }
\label{fig::frequency}
\end{figure}

 We turn our attention to the case where $I_{deg}$ and $I_{clos}$ do not intersect. In this case, we expect from theory a transition in the nodes that belong to $I_{dc}^{\delta}$ as $\delta$ increases. It turns out that most of the times the transition is immediate, meaning that any node in $I_{dc}^{\delta}$ belongs to either $I_{deg}$ or $I_{clos}$. This can become apparent in Figure~\ref{fig::frequency}, which contains the percentage frequencies with which $I_{dec}^{\delta} \subseteq I_{deg}$, $I_{dec}^{\delta} \subseteq I_{clos}$ and $I_{dc}^{\delta} \cap \left(I_{deg}\cup I_{clos}\right)=\emptyset$. For each value of $\delta$ the frequencies correspond to the fraction of the simulated networks in which each of the three conditions held true. The fact that in the left subfigure $I_{deg}$ and $I_{clos}$ can intersect means that the sum of the frequencies may exceed 100\%, which seems to be the case rather often. This is no longer possible in the right subfigure, where we include only the cases where $I_{deg}$ and $I_{clos}$ do not intersect, therefore the three conditions are mutually exclusive. Note that, observing a node that maximizes decay centrality without belonging either to $I_{deg}$ or $I_{clos}$ is never observed in more than a 2\% of the trials, with the percentage becoming much lower away from $\delta=0.5$. This suggests that for most networks there is a threshold value of $\delta$ below which $I_{dec}^{\delta} \subseteq I_{deg}$ and above which $I_{dec}^{\delta} \subseteq I_{clos}$.
 
\begin{figure}[htbp]
\vspace{-1cm}
\centering
\makebox[\linewidth][c]{%
\begin{subfigure}[b]{.58\textwidth}
        \centering
        \resizebox{\linewidth}{!}{
\includegraphics{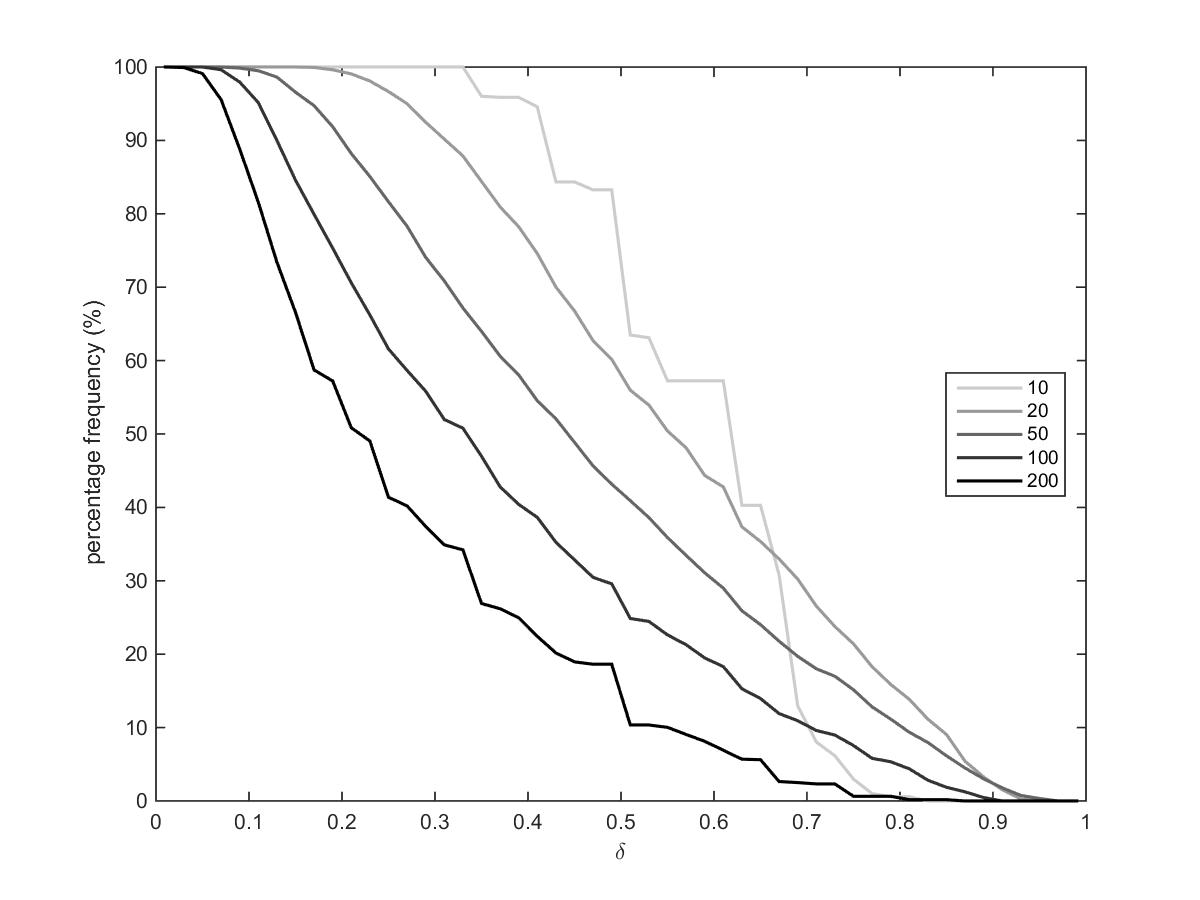}
}
\end{subfigure}
    \begin{subfigure}[b]{.58\textwidth}
        \centering
        \resizebox{\linewidth}{!}{
\includegraphics{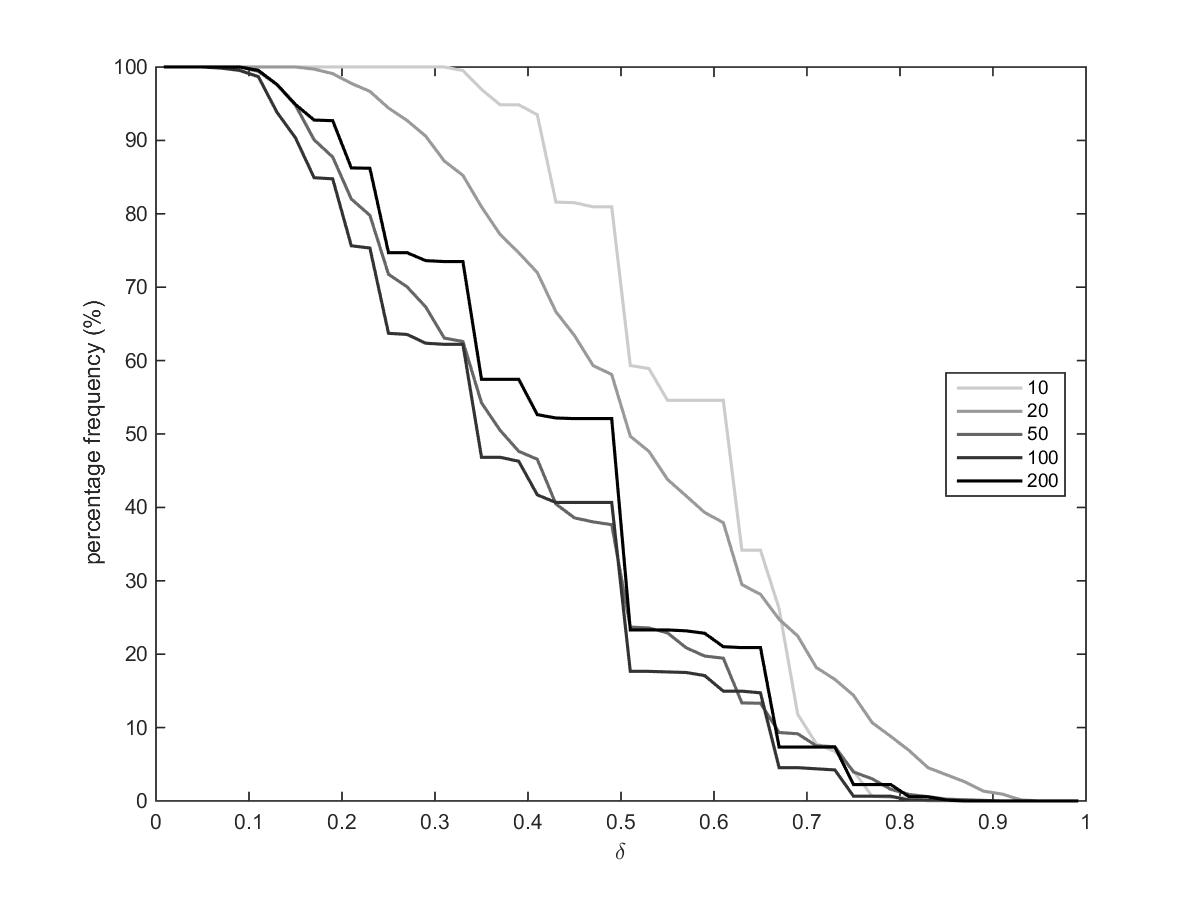}
}
\end{subfigure}
}\\
\vspace{-0.3cm}
\makebox[\linewidth][c]{%
    \begin{subfigure}[b]{.58\textwidth}
        \centering
        \resizebox{\linewidth}{!}{
\includegraphics{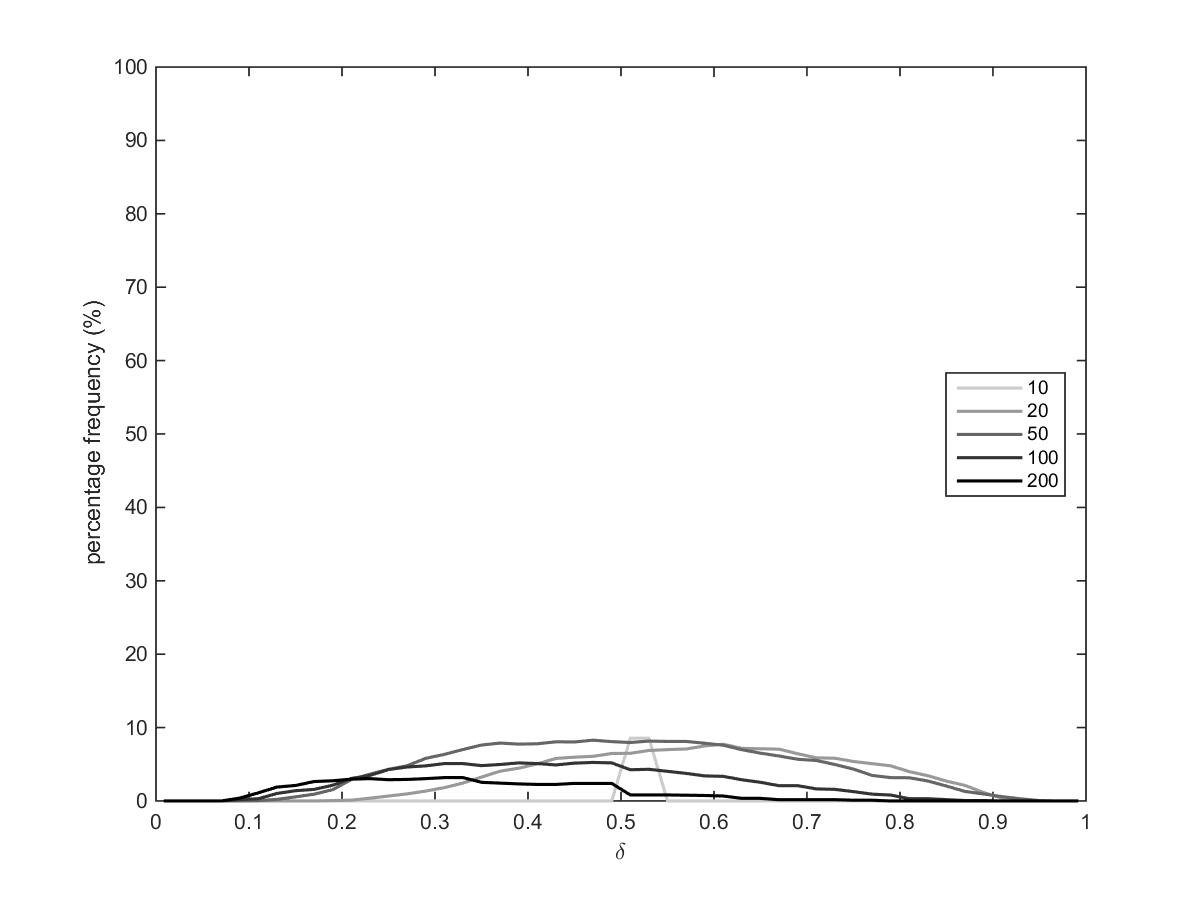}
}
\end{subfigure}
\begin{subfigure}[b]{.58\textwidth}
        \centering
        \resizebox{\linewidth}{!}{
\includegraphics{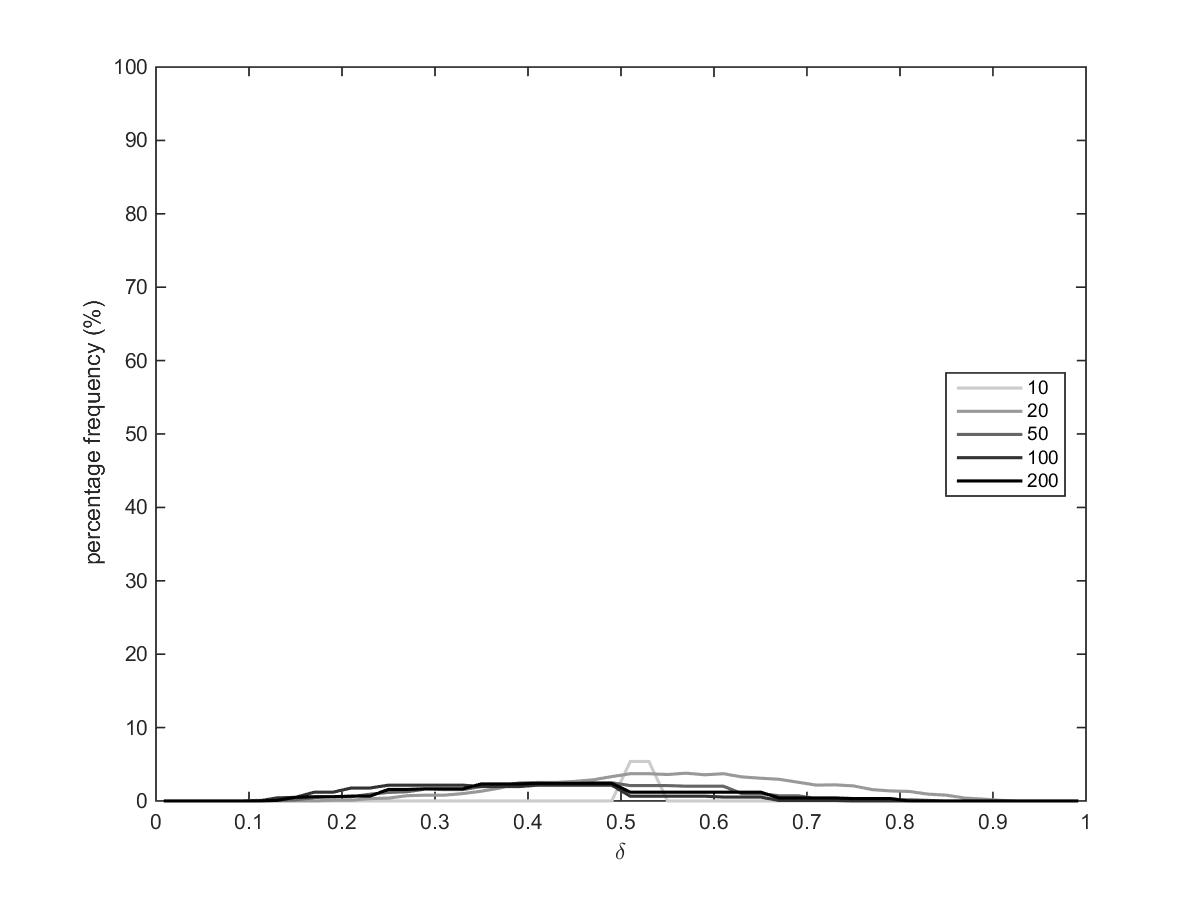}
}
\end{subfigure}
}\\
\vspace{-0.3cm}
\makebox[\linewidth][c]{%
    \begin{subfigure}[b]{.58\textwidth}
        \centering
        \resizebox{\linewidth}{!}{
\includegraphics{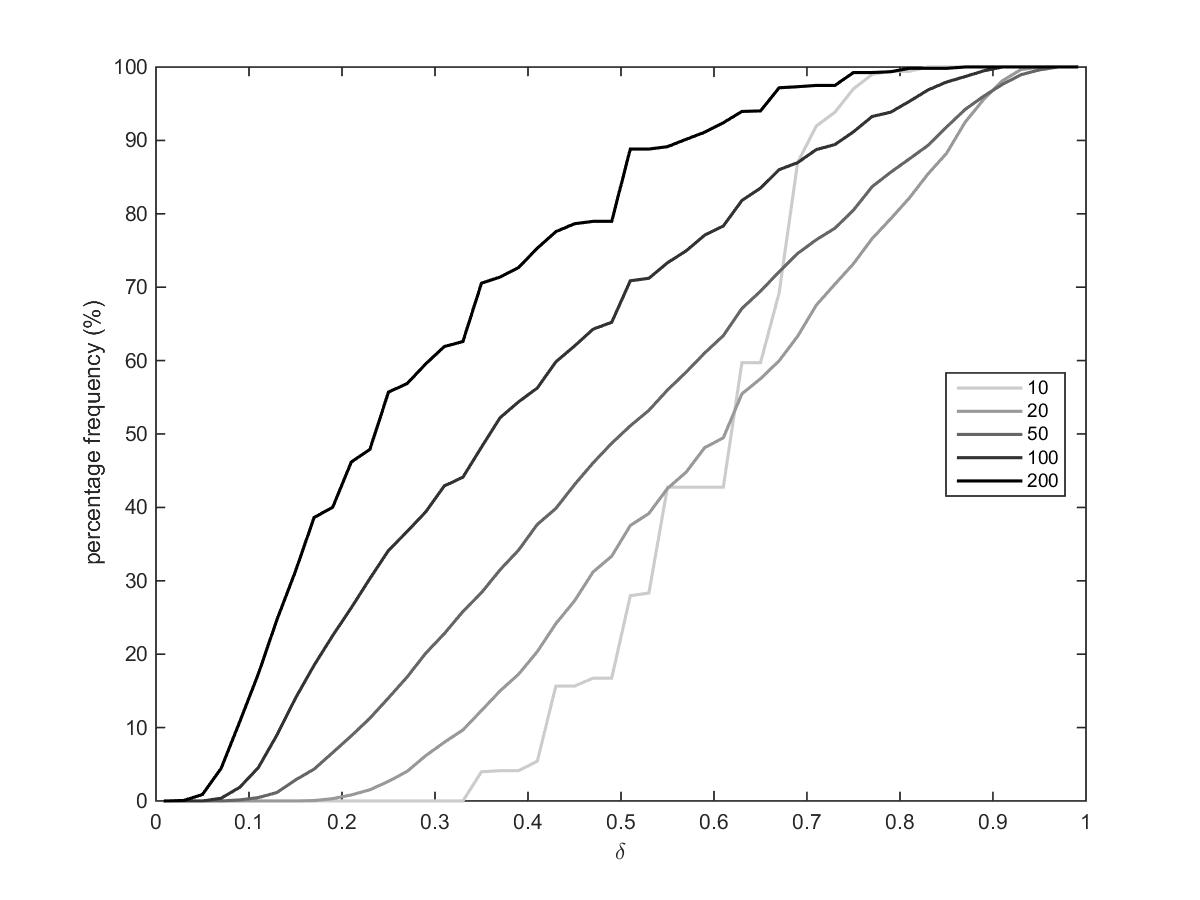}
}
\end{subfigure}
\begin{subfigure}[b]{.58\textwidth}
        \centering
        \resizebox{\linewidth}{!}{
\includegraphics{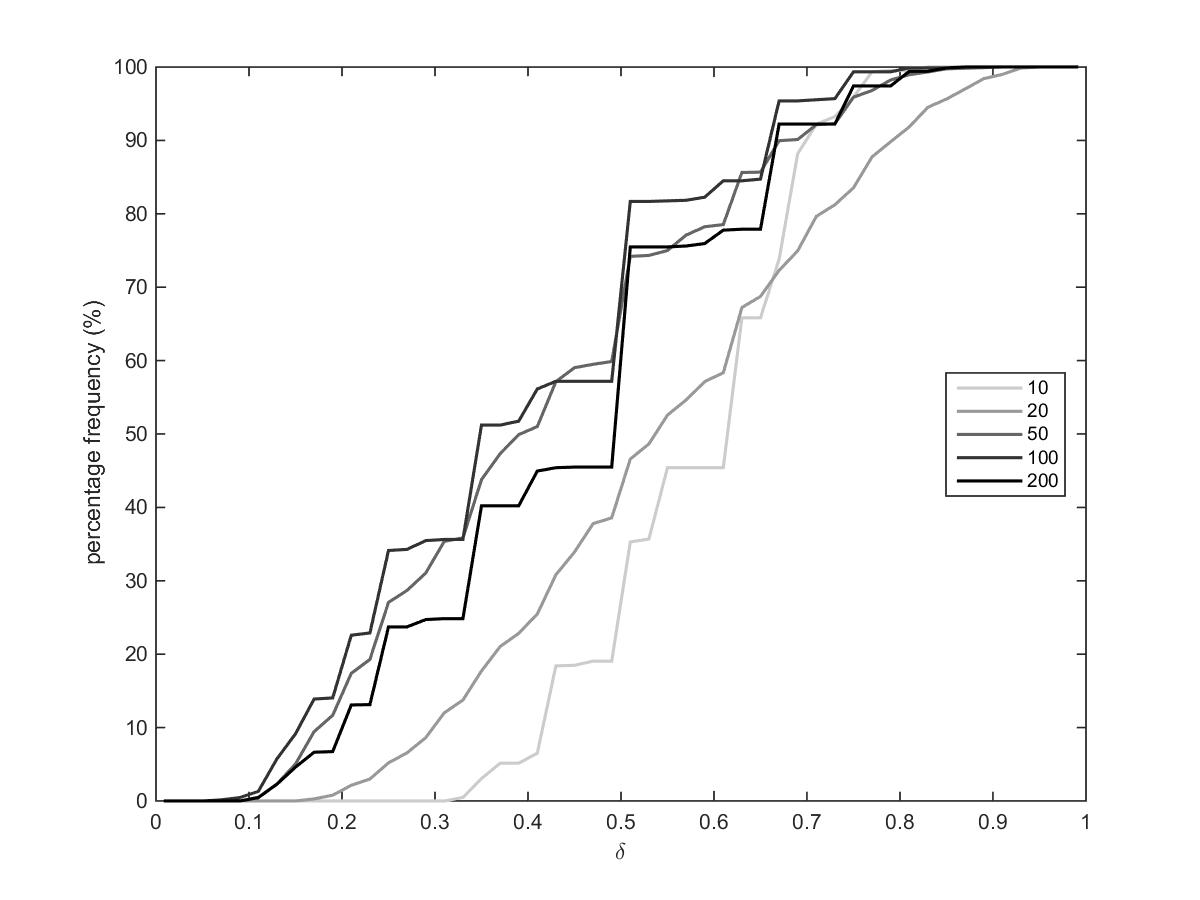}
}
\end{subfigure}
}
\caption{Percentage frequency of networks for which $I_{dec}^{\delta} \subseteq I_{deg}$ (first row), $I_{dc}^{\delta} \cap \left(I_{deg}\cup I_{clos}\right)=\emptyset$ (second row) and $I_{dec}^{\delta} \subseteq I_{clos}$ (third row) for each value of the decay parameter, presented separately for each network size. The two columns correspond to $p=0.05$ (left) and $p=0.1$ (right).}
\label{fig::histpernet_prob05_to10}
\end{figure}

\begin{figure}[htbp]
\vspace{-1cm}
\centering
\makebox[\linewidth][c]{%
\begin{subfigure}[b]{.58\textwidth}
        \centering
        \resizebox{\linewidth}{!}{
\includegraphics{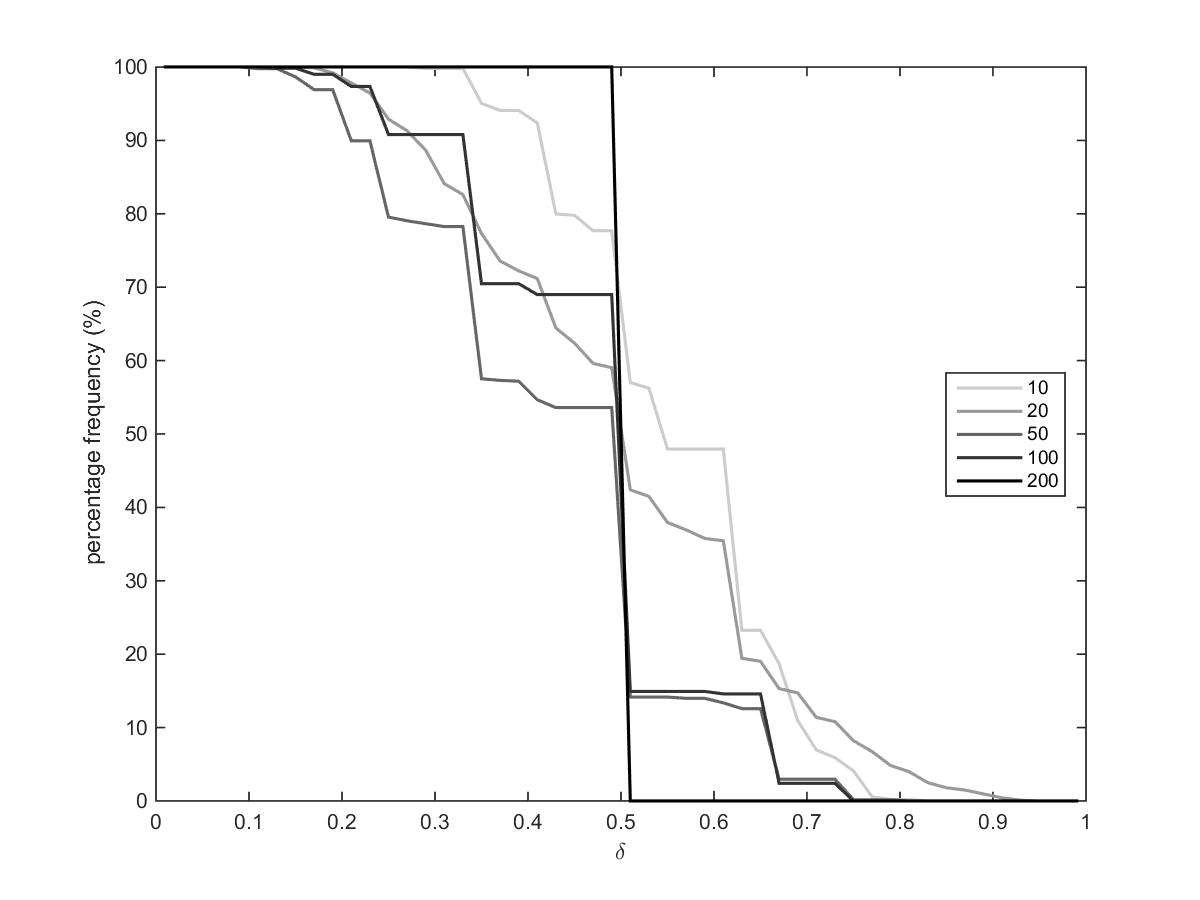}
}
\end{subfigure}
    \begin{subfigure}[b]{.58\textwidth}
        \centering
        \resizebox{\linewidth}{!}{
\includegraphics{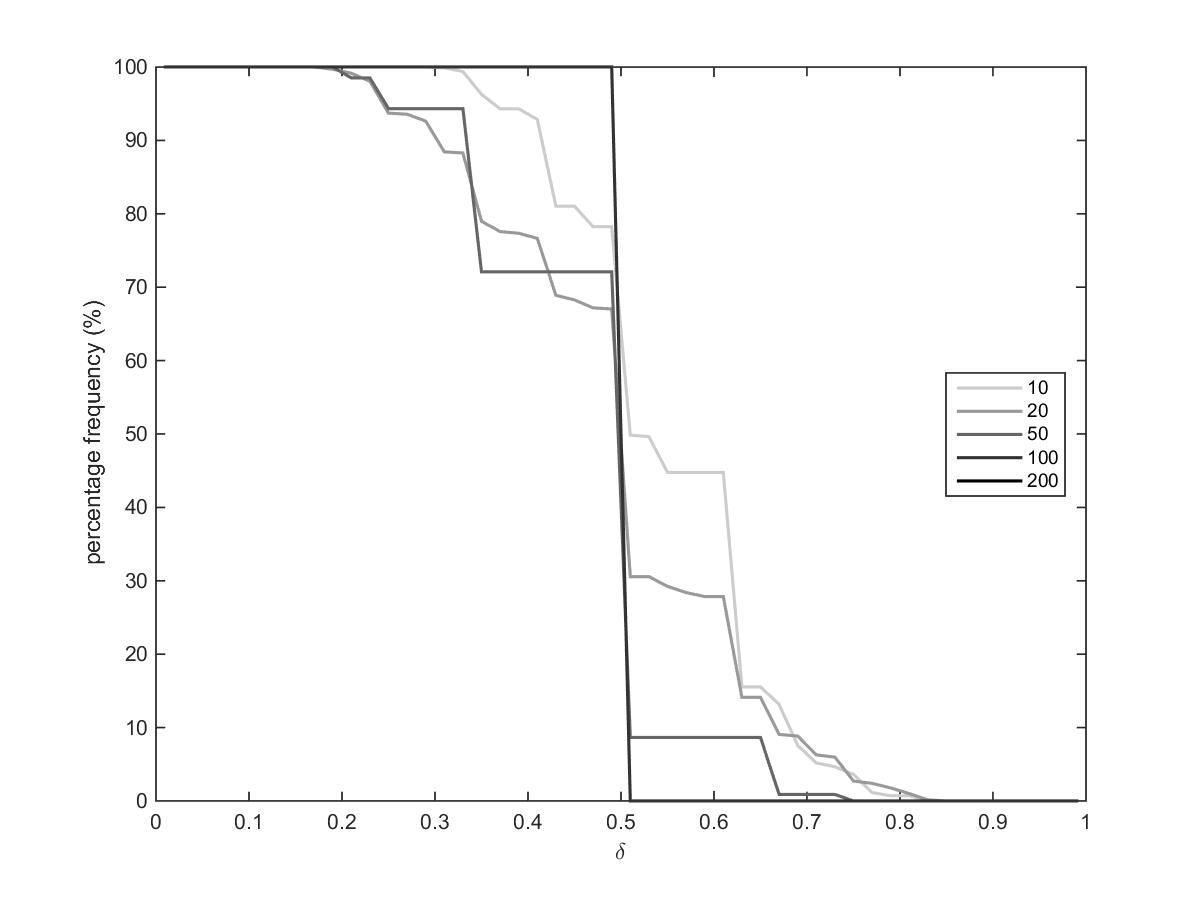}
}
\end{subfigure}
}\\
\vspace{-0.3cm}
\makebox[\linewidth][c]{%
    \begin{subfigure}[b]{.58\textwidth}
        \centering
        \resizebox{\linewidth}{!}{
\includegraphics{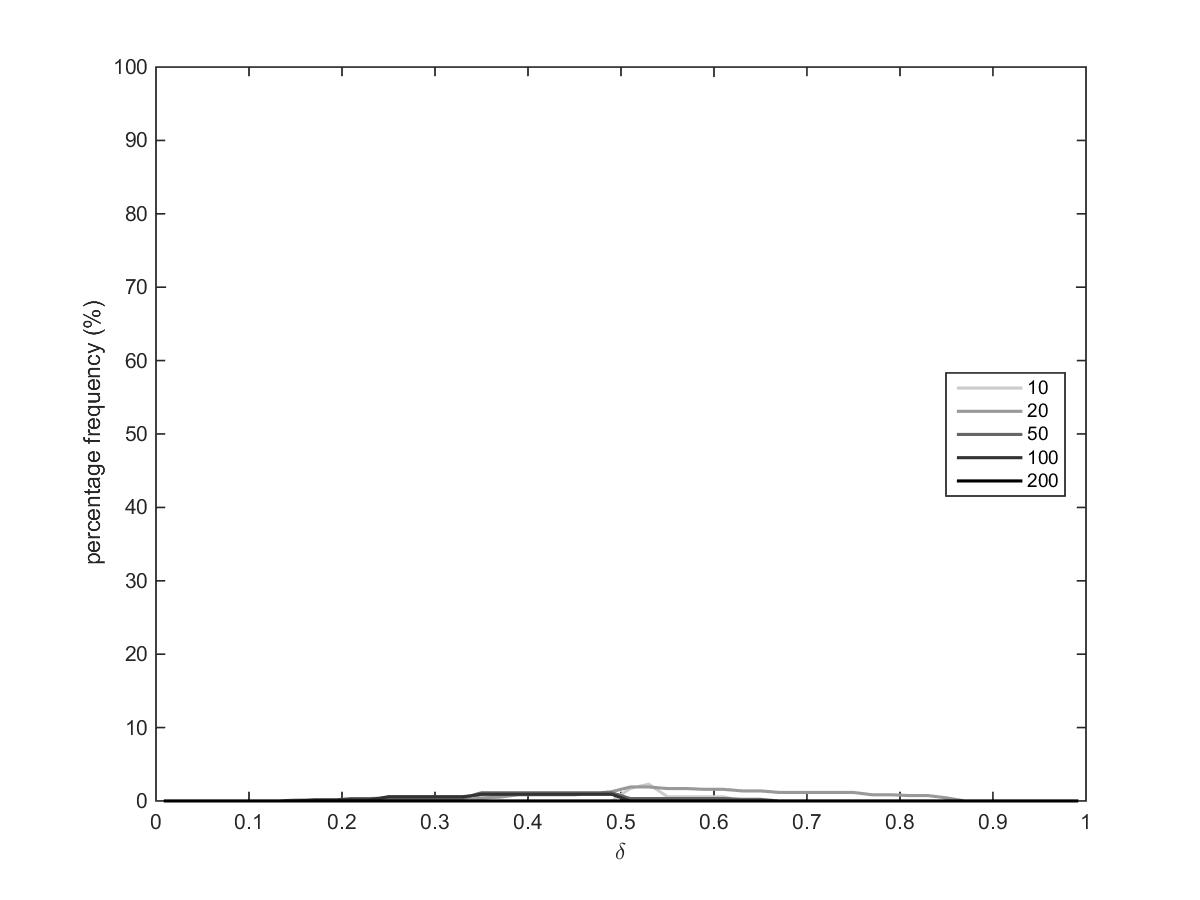}
}
\end{subfigure}
\begin{subfigure}[b]{.58\textwidth}
        \centering
        \resizebox{\linewidth}{!}{
\includegraphics{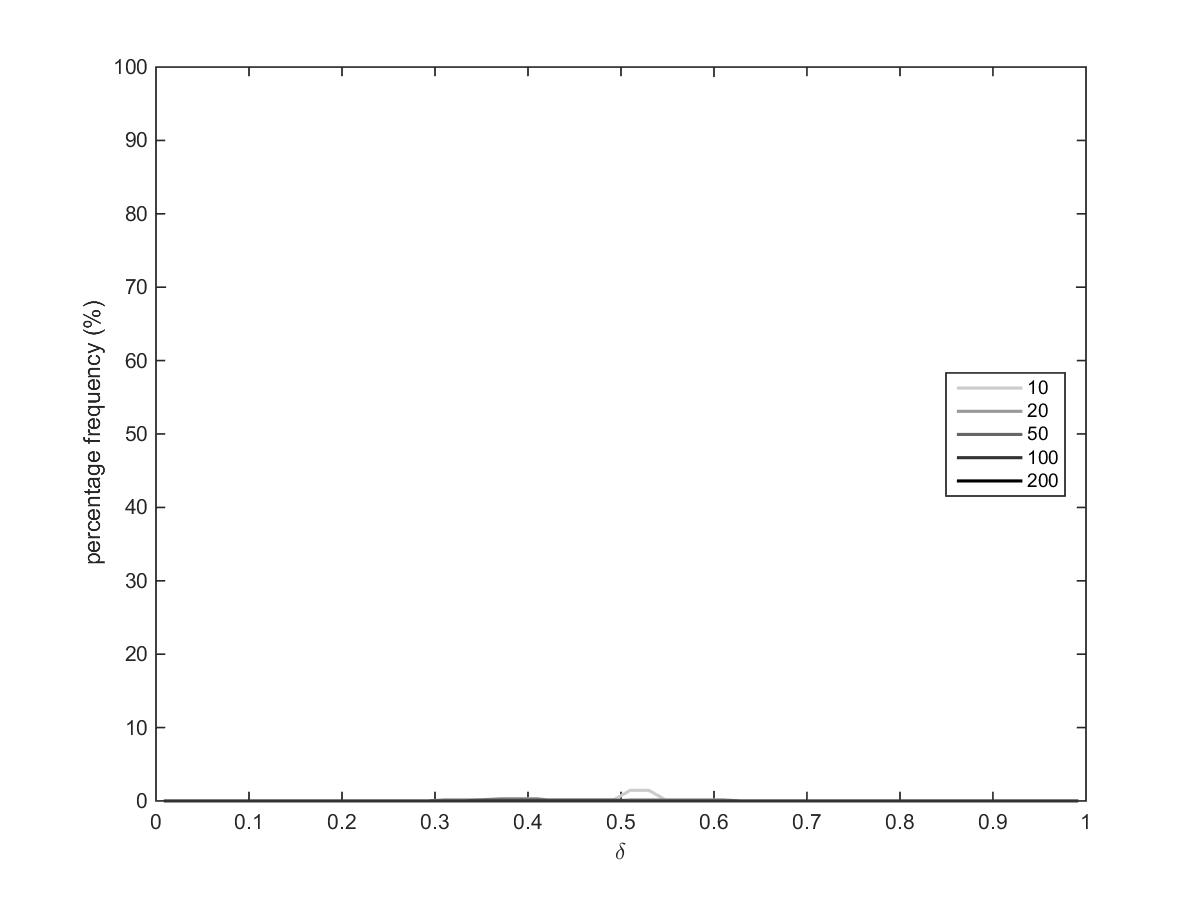}
}
\end{subfigure}
}\\
\vspace{-0.3cm}
\makebox[\linewidth][c]{%
    \begin{subfigure}[b]{.58\textwidth}
        \centering
        \resizebox{\linewidth}{!}{
\includegraphics{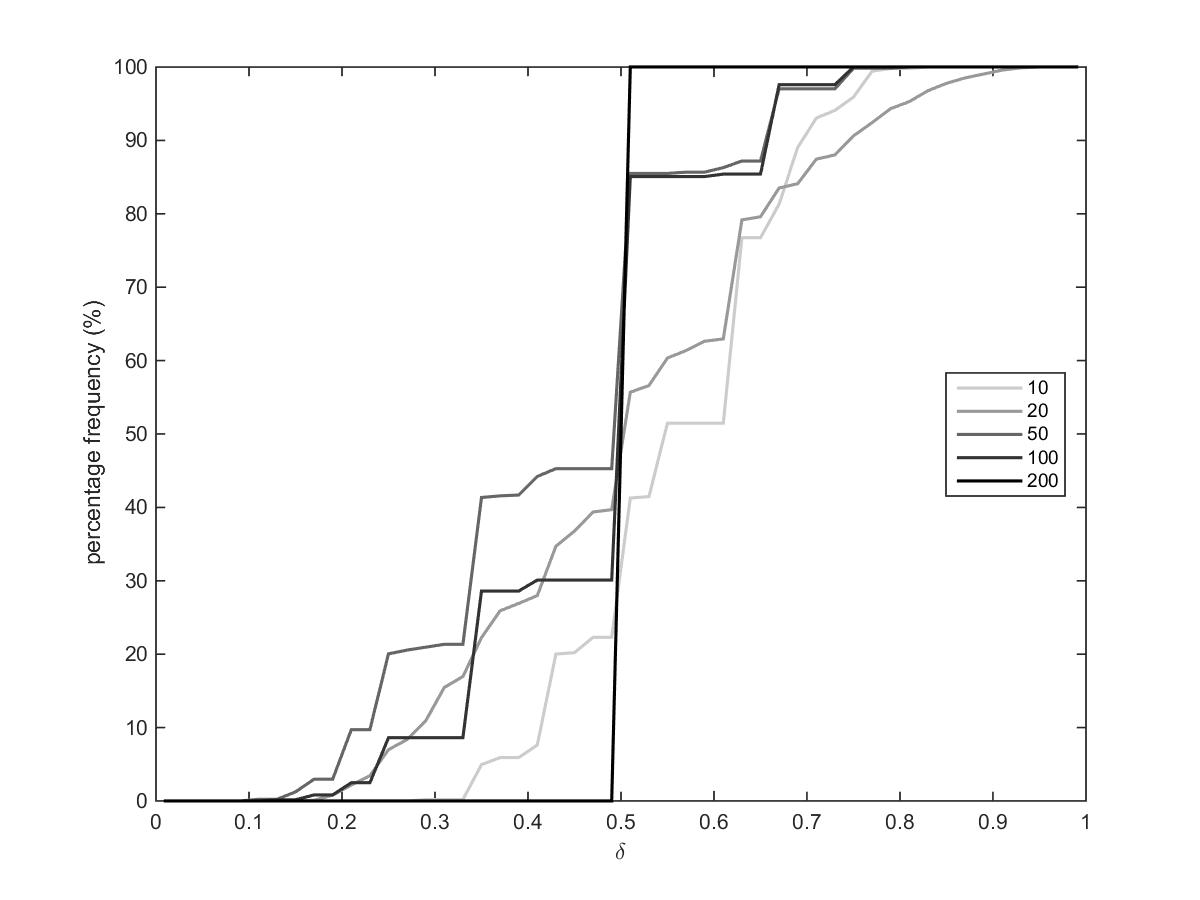}
}
\end{subfigure}
\begin{subfigure}[b]{.58\textwidth}
        \centering
        \resizebox{\linewidth}{!}{
\includegraphics{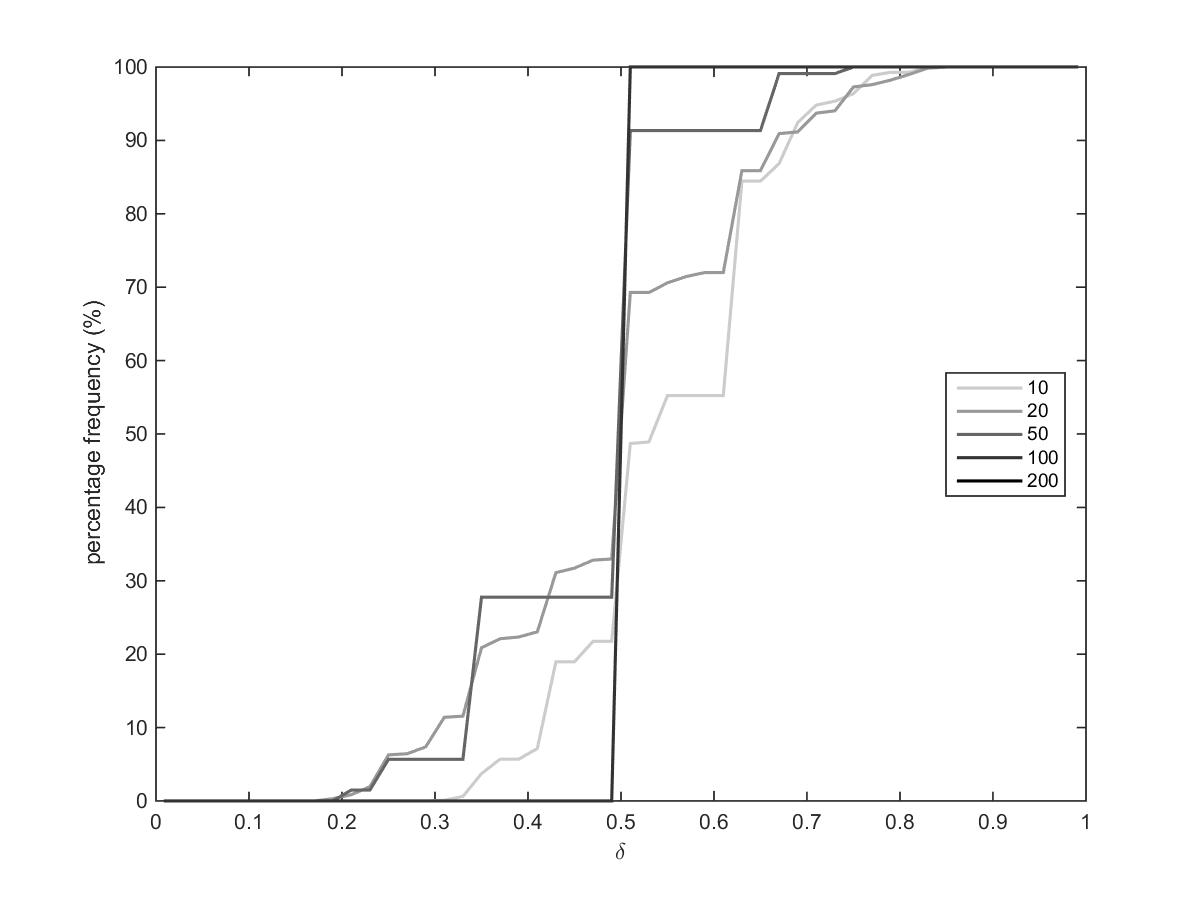}
}
\end{subfigure}
}
\caption{Percentage frequency of networks for which $I_{dec}^{\delta} \subseteq I_{deg}$ (first row), $I_{dc}^{\delta} \cap \left(I_{deg}\cup I_{clos}\right)=\emptyset$ (second row) and $I_{dec}^{\delta} \subseteq I_{clos}$ (third row) for each value of the decay parameter, presented separately for each network size. The two columns correspond to $p=0.15$ (left) and $p=0.2$ (right).}
\label{fig::histpernet_prob15_to20}
\end{figure}

Figures~\ref{fig::histpernet_prob05_to10} and \ref{fig::histpernet_prob15_to20} contain the three percentage frequencies of interest for all network sizes and values of $p \in \{0.05, 0.1, 0.15, 0.2\}$, focusing on networks where $I_{deg}$ and $I_{clos}$ do not intersect. The results are qualitatively similar in all configurations, presenting an inverted S-shaped curve for the frequency of $I_{dec}^{\delta} \subseteq I_{deg}$, an S-shaped curve for $I_{dec}^{\delta} \subseteq I_{clos}$ and an inverted bell curve for $I_{dc}^{\delta} \cap \left(I_{deg}\cup I_{clos}\right)=\emptyset$. 

\noindent The latter one never exceeds 10\%, with this being the case only for $p=0.05$ and values of $\delta$ close to 0.5. We also observe the transition of maximizers of decay centrality from belonging to the set of nodes with maximum degree to that of nodes with maximum closeness occurs for lower values of $\delta$ as networks become larger, with the result being more prevalent for low values of $p$; as $p$ increases we observe a sharp transition occurring for $\delta$ around 0.5.\footnote{For $p\geq0.2$ there are very few observations where $I_{deg}$ and $I_{clos}$ do not intersect.}

\begin{figure}[htbp]
\vspace{-1.1cm}
\centering
    \begin{subfigure}[b]{0.63\textwidth}
        \centering
        \resizebox{\linewidth}{!}{
\includegraphics{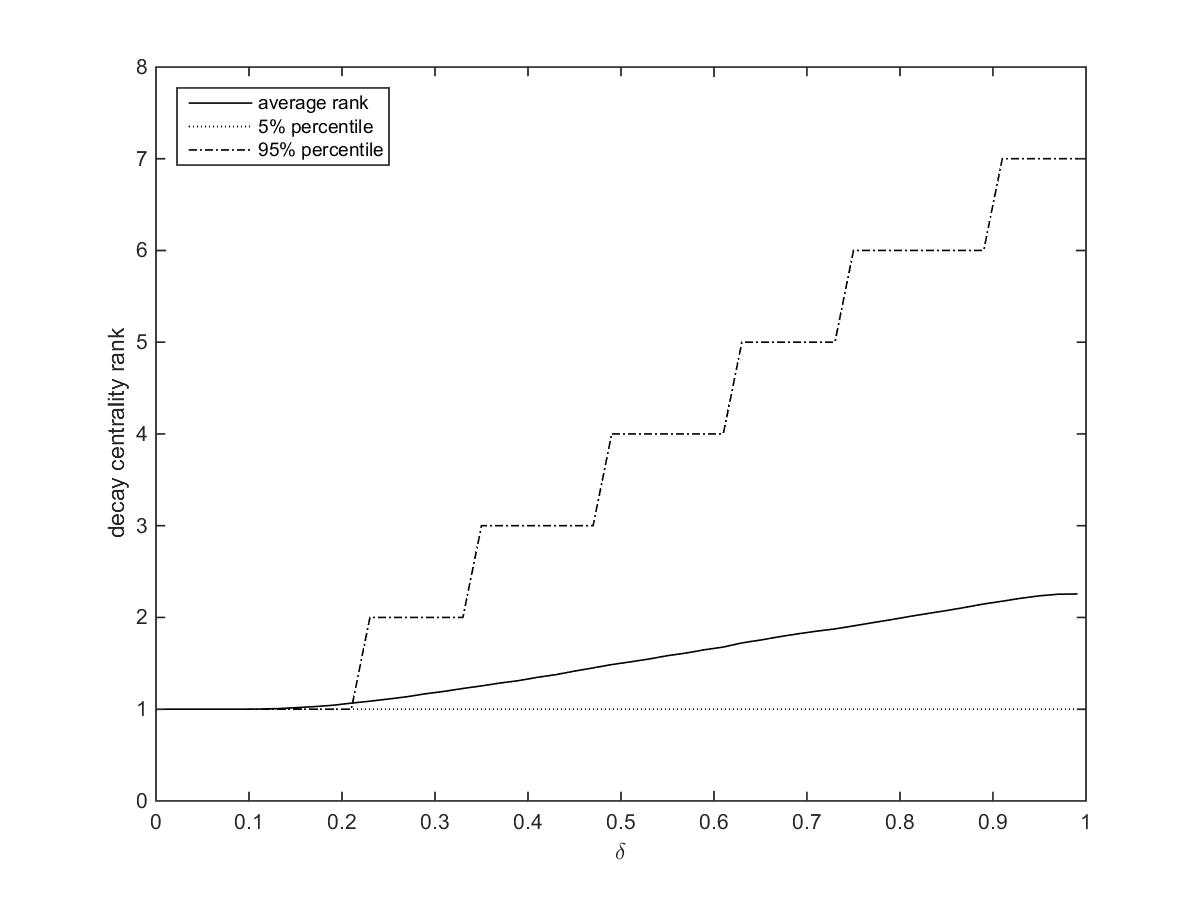}
}
\end{subfigure}
\centering
    \begin{subfigure}[b]{0.63\textwidth}
    \vspace{-0.6cm}
        \centering
        \resizebox{\linewidth}{!}{
\includegraphics{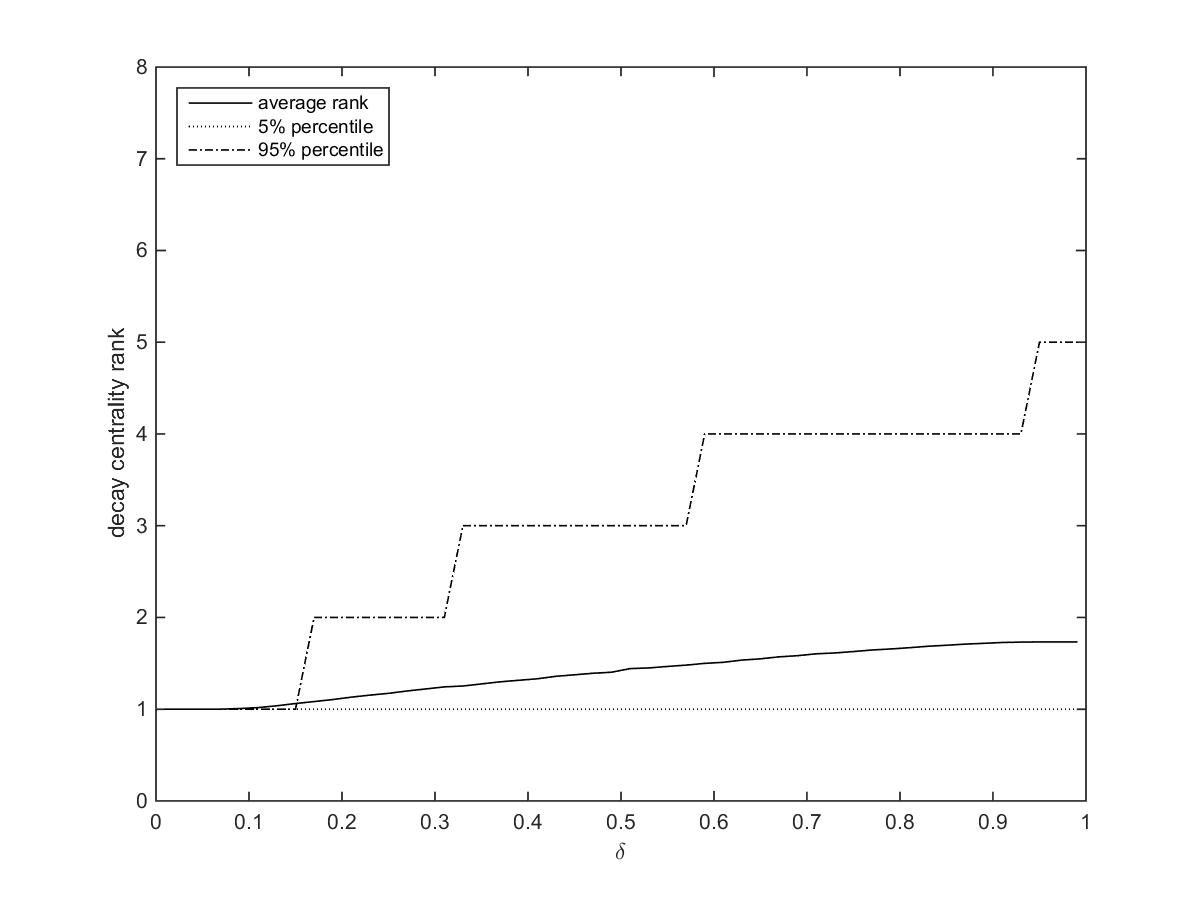}
}
\end{subfigure}
    \begin{subfigure}[b]{0.63\textwidth}
    \vspace{-0.6cm}
        \centering
        \resizebox{\linewidth}{!}{
\includegraphics{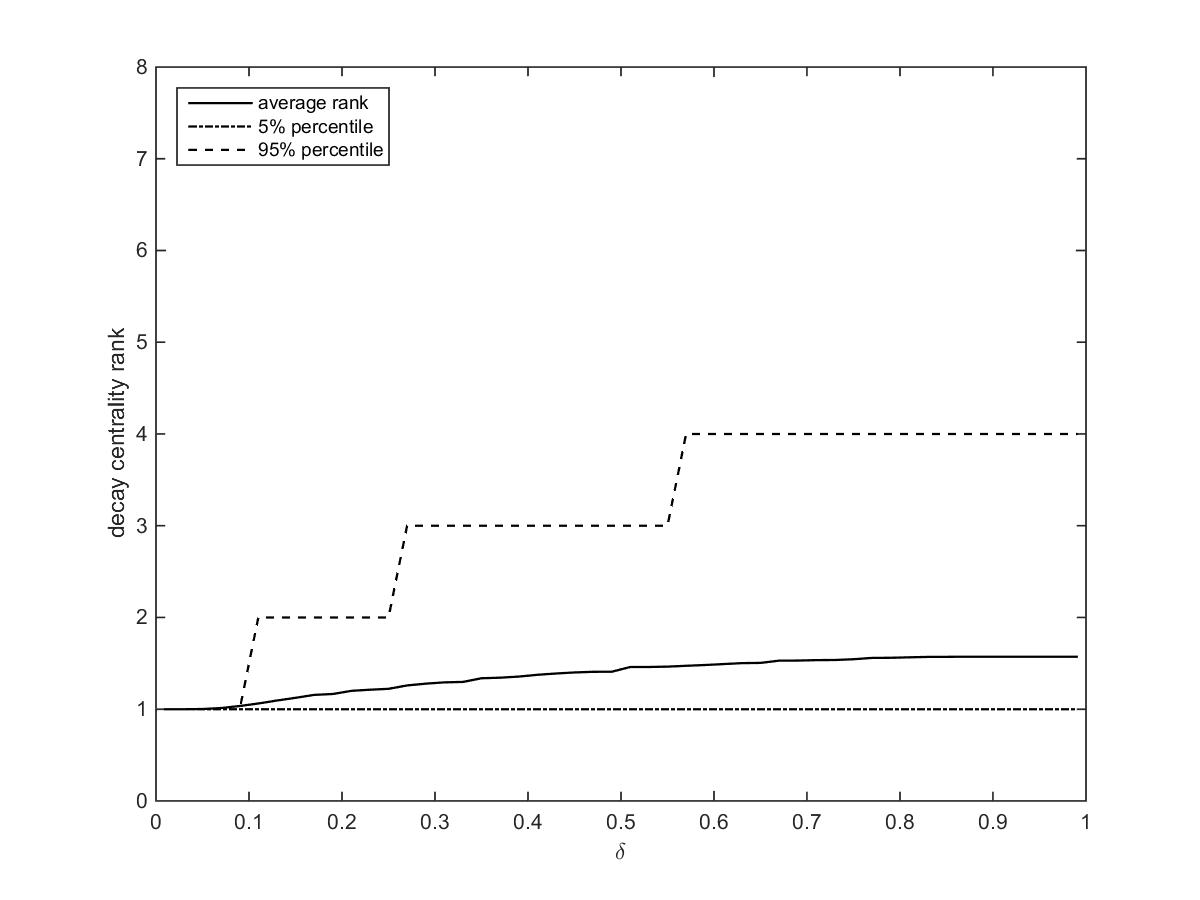}
}
\end{subfigure}
    \vspace{-0.3cm}
\caption{Average rank (solid), 5th (dotted) and 95th (dashed) percentiles of decay centrality of nodes with maximum degree. Subfigures correspond to $n=50, 100, 200$ from top to bottom and $p=0.05$.}
\label{fig::rank_degree_p05_allnets}
\end{figure}

\begin{figure}[htbp]
\vspace{-1.1cm}
\centering
    \begin{subfigure}[b]{0.63\textwidth}
        \centering
        \resizebox{\linewidth}{!}{
\includegraphics{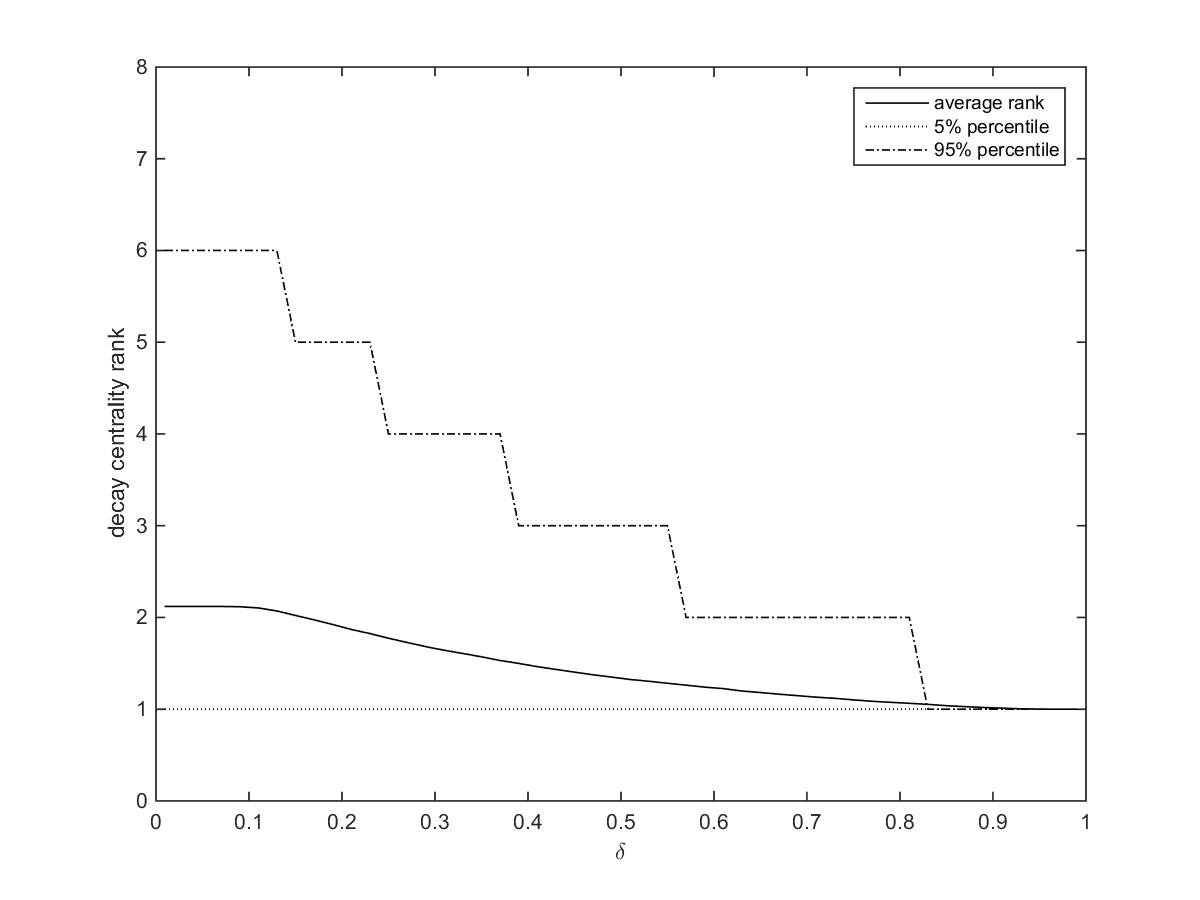}
}
\end{subfigure}
\centering
    \begin{subfigure}[b]{0.63\textwidth}
    \vspace{-0.6cm}
        \centering
        \resizebox{\linewidth}{!}{
\includegraphics{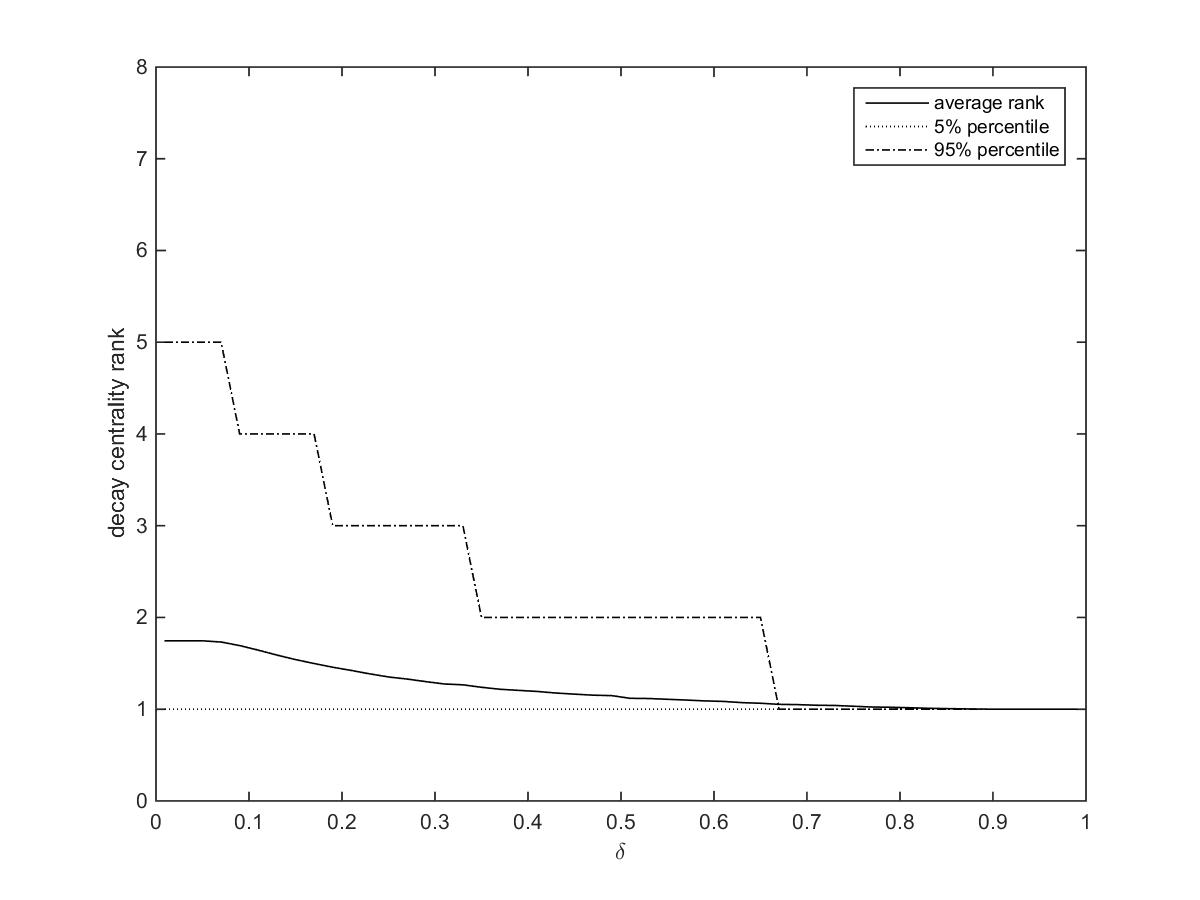}
}
\end{subfigure}
    \begin{subfigure}[b]{0.63\textwidth}
    \vspace{-0.6cm}
        \centering
        \resizebox{\linewidth}{!}{
\includegraphics{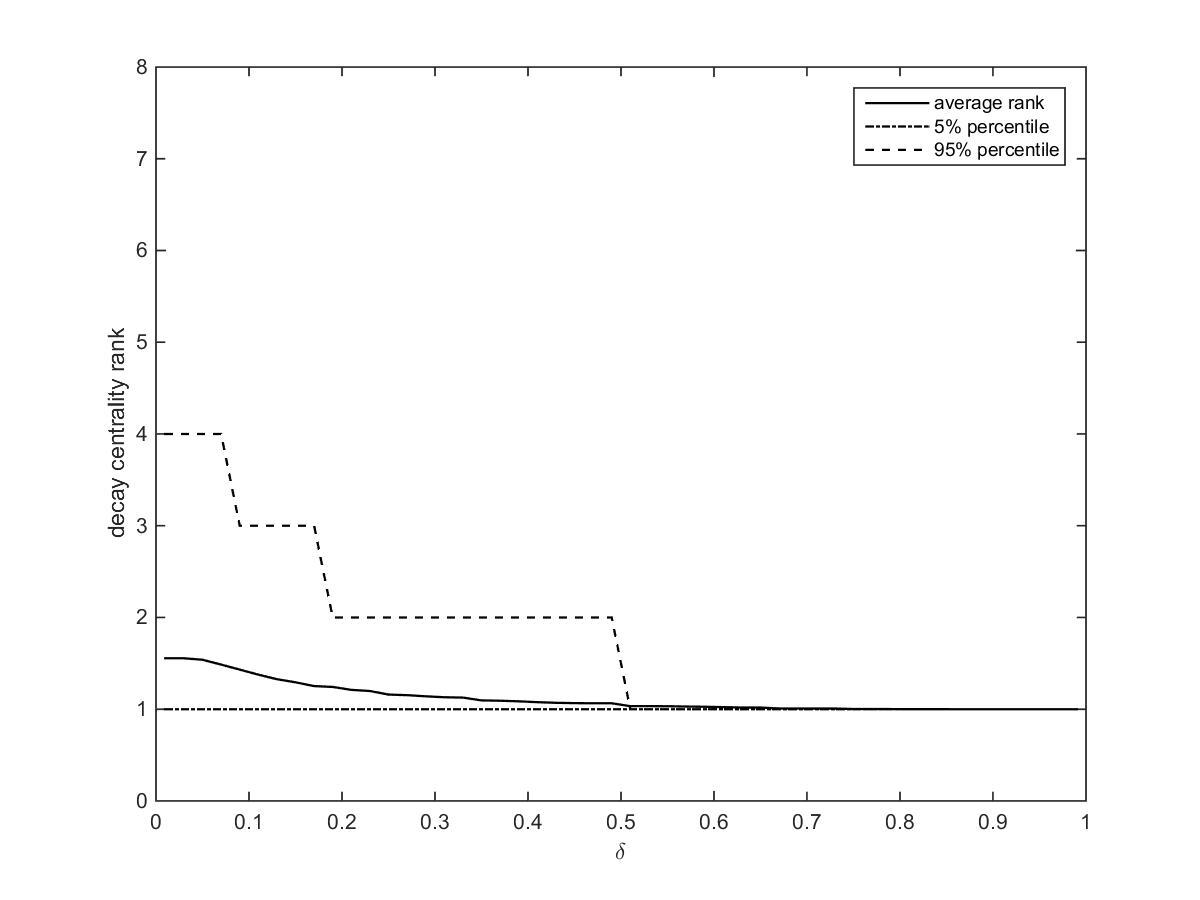}
}
\end{subfigure}
    \vspace{-0.3cm}
\caption{Average rank (solid), 5th (dotted) and 95th (dashed) percentiles of decay centrality of nodes with maximum closeness. Subfigures correspond to $n=50, 100, 200$ from top to bottom and $p=0.05$.}
\label{fig::rank_closeness_p05_allnets}
\end{figure}

At this point one might wonder what is the rank in terms of decay centrality of nodes with maximum degree and closeness, when they are not ranked first? Figures~\ref{fig::rank_degree_p05_allnets} and~\ref{fig::rank_closeness_p05_allnets} show the average rank, as well as the 5th and 95th percentiles of rank distribution in decay centrality of nodes with maximum degree and closeness respectively. It turns out that nodes belonging to $I_{deg}$ or $I_{clos}$ are highly ranked in terms of decay centrality for all values of $\delta$, even when they are not ranked first. The result is similar if we exclude the networks in which $I_{deg}$ and $I_{clos}$ intersect. 

\begin{figure}[htbp]
\vspace{-1.2cm}
\centering
    \begin{subfigure}[b]{0.625\textwidth}
        \centering
        \resizebox{\linewidth}{!}{
\includegraphics{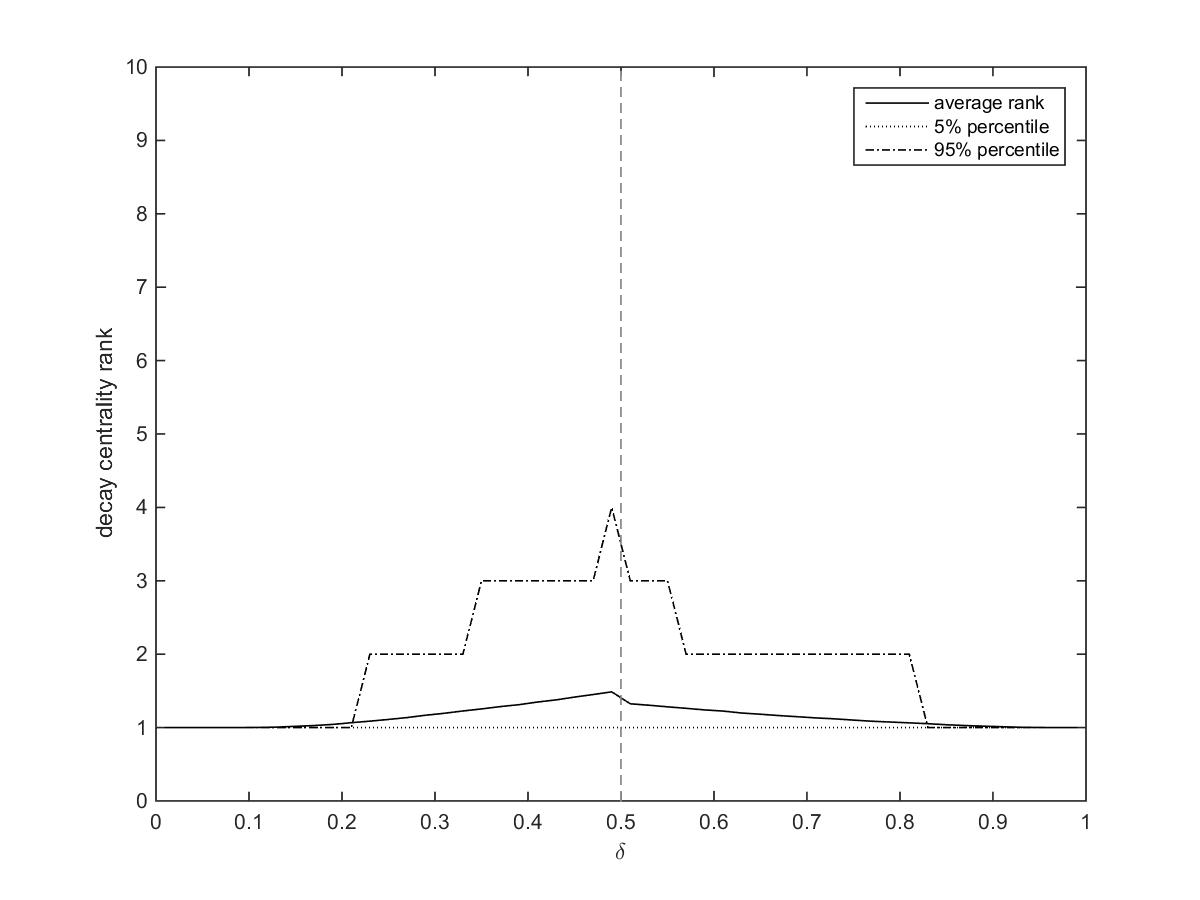}
}
\end{subfigure}
\centering
    \begin{subfigure}[b]{0.625\textwidth}
    \vspace{-0.65cm}
        \centering
        \resizebox{\linewidth}{!}{
\includegraphics{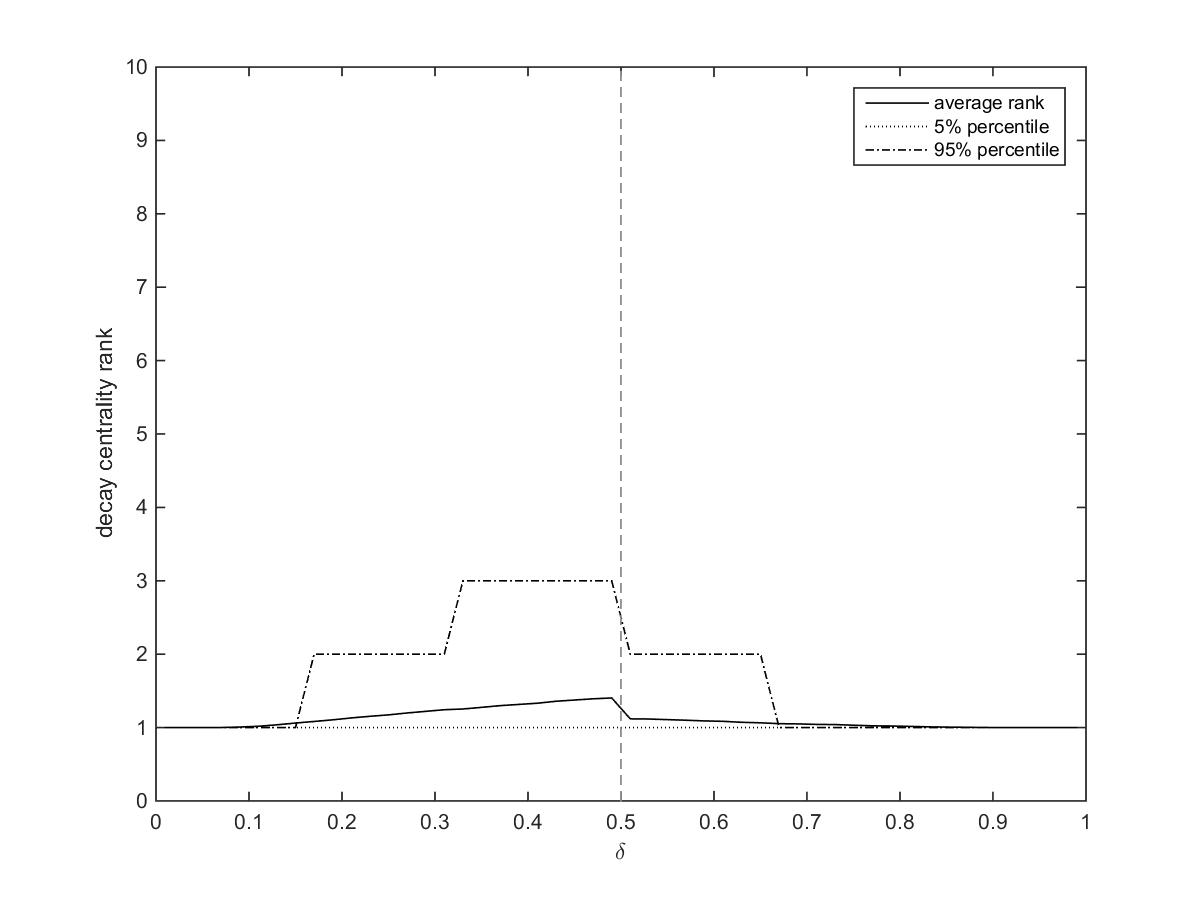}
}
\end{subfigure}
    \begin{subfigure}[b]{0.625\textwidth}
    \vspace{-0.65cm}
        \centering
        \resizebox{\linewidth}{!}{
\includegraphics{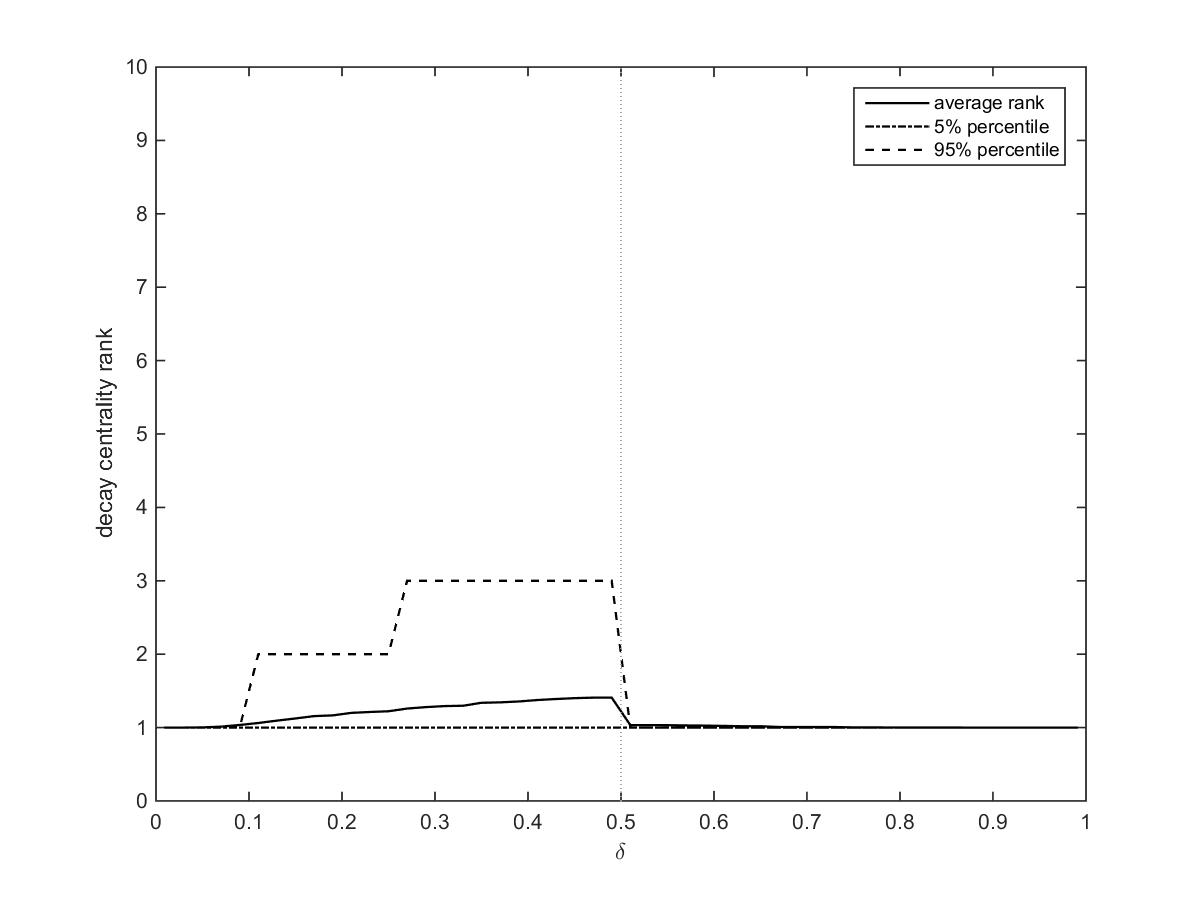}
}
\end{subfigure}
    \vspace{-0.4cm}
\caption{Average rank (solid), 5th (dotted) and 95th (dashed) percentiles, of decay centrality of nodes with maximum degree for $\delta<0.5$ and maximum closeness for $\delta>0.5$. Subfigures correspond to $n=50, 100, 200$ from top to bottom and $p=0.05$.}
\label{fig::rank_combined_p05_allnets}
\end{figure}

Nevertheless, focusing only on degree or only on closeness leads to increasingly suboptimal choices as $\delta$ moves towards the extremes. Hence, it remains unclear which of the two sets would provide better candidates depending on the value of $\delta$. Ideally, we would like to have a simple rule of thumb that would facilitate this choice. A natural rule would be to choose a node from $I_{deg}$ for $\delta<0.5$ and a node from $I_{clos}$ for $\delta>0.5$. Given that the two sets usually contain few nodes, it should not be too costly to calculate the decay centrality of each of these nodes and pick the one that maximizes it. It turns out that this rule of thumb is sufficient to ensure that the chosen node will be highly ranked in terms of decay centrality. Figure~\ref{fig::rank_combined_p05_allnets} shows the same statistics as Figures \ref{fig::rank_degree_p05_allnets} and \ref{fig::rank_closeness_p05_allnets} for $p=0.05$ and all network sizes, in which it can be seen that in all networks a node chosen according to this rule will be ranked in terms of decay centrality among the top three with probability 95\%. Additional figures in the Online Appendix suggest that the result is robust to different parameter values and even for networks with 1000 nodes.

\section{Discussion}

We have established, both analytically and numerically, a relation between decay centrality, degree and closeness, showing that nodes that maximize one of the two measures are natural candidates for maximizing decay centrality as well. In fact, the majority of networks has a threshold value of $\delta$ below which maximum decay centrality coincides with maximum degree and above which it coincides with maximum closeness. We show that a simple rule of thumb that considers a common threshold at $\delta=0.5$ performs particularly well. It still remains rather unexplored whether there are particular characteristics of the network that can allow the accurate characterization of this threshold value with some accuracy. Finally, simulations are limited to networks of small to medium size. In the Online Appendix we provide some additional support in favor of our results by considering networks with 1000 nodes, however a more systematic extension of the analysis to large networks would ensure their applicability to problems where decay centrality has been shown to play an important role.

\begin{appendix}

\section{Proofs}

\begin{proof}[\textbf{\textup{Proof of Proposition~\ref{proplowdelta}}}]
Notice that decay centrality can be rewritten as $DC_i(\delta)=\delta D_i+\sum\limits_{l=2}^{n-1}\delta^lD_i^l$ and consider the limit $\lim\limits_{\delta \to 0} \frac{DC_i(\delta)-DC_j(\delta)}{\delta}\overset{\frac{0}{0}}{=}D_i-D_j>0$, where the last inequality is implied by $D_i>D_j$. The expression that appears inside the limit is continuous in $\delta$, therefore exists $\underline{\delta}$ such that for all $\delta \in (0,\underline{\delta})$ it holds that $\frac{DC_i(\delta)-DC_j(\delta)}{\delta}>0$ or equivalently $DC_i(\delta)>DC_j(\delta)$.
\end{proof}

\begin{proof}[\textbf{\textup{Proof of Proposition~\ref{proplowdeltageneral}}}]
If $D_i>D_j$ then the result holds immediately by Proposition \ref{proplowdelta}. Otherwise, let $\tilde{l}\geq2$ be the first instance such that $D_i^{\tilde{l}}>D_j^{\tilde{l}}$. In this case, the difference between decay centralities can be written as $DC_i(\delta)-DC_j(\delta)=\sum\limits_{l=\tilde{l}}^{n-1}\delta^l(D_i^l-D_j^l)$, because all previous terms of the sum are equal to zero. Hence, $\lim\limits_{\delta \to 0} \frac{DC_i(\delta)-DC_j(\delta)}{\delta^{\tilde{l}}}\overset{\frac{0}{0}}{=}D_i^{\tilde{l}}-D_j^{\tilde{l}}>0$ and by continuity with respect to $\delta$, there is $\underline{\delta}$ such that for all $\delta \in (0,\underline{\delta})$ it holds that $\frac{DC_i(\delta)-DC_j(\delta)}{\delta^{\tilde{l}}}>0$ or equivalently $DC_i(\delta)>DC_j(\delta)$.
\end{proof}

\begin{proof}[\textbf{\textup{Proof of Proposition~\ref{prophighdelta}}}]
We need three direct observations to obtain the result. First, note that $C_i>C_j \Leftrightarrow \sum\limits_{k \neq i} d(i,k)<\sum\limits_{k \neq j} d(j,k)$. Second, recall the alternative formulation of decay centrality as $DC_i(\delta)=\delta D_i+\sum\limits_{l=2}^{n-1}\delta^lD_i^l$ and third to observe that $\sum\limits_{k \neq i} d(i,k)=D_i+\sum\limits_{l=2}^{n-1}lD_i^l$. Therefore, 
\begin{align*}
\lim\limits_{\delta \to 1}\frac{DC_i(\delta)-DC_j(\delta)}{1-\delta}&=\lim\limits_{\delta \to 1}\frac{\left(\delta D_i+\sum\limits_{l=2}^{n-1}\delta^lD_i^l\right)-\left(\delta D_j+\sum\limits_{l=2}^{n-1}\delta^lD_j^l\right)}{1-\delta}\overset{\frac{0}{0}}{=}\\
&=\lim\limits_{\delta \to 1}\frac{\left(D_i+\sum\limits_{l=2}^{n-1}l\delta^{l-1}D_i^l\right)-\left(D_j+\sum\limits_{l=2}^{n-1}l\delta^{l-1}D_j^l\right)}{-1}=\\
&=-\left(\sum\limits_{k\neq i}d(i,k)-\sum\limits_{k\neq j}d(j,k)\right)>0
\end{align*}
where the last inequality is implied by $C_i>C_j$. Hence, by continuity with respect to $\delta$, exists $\overline{\delta}$ such that for all $\delta \in (\overline{\delta},1)$ holds that $\frac{DC_i(\delta)-DC_j(\delta)}{1-\delta}>0$ or equivalently $DC_i(\delta)>DC_j(\delta)$.
\end{proof}

\begin{proof}[\textbf{\textup{Proof of Proposition~\ref{prophighdeltageneral}}}]
If $C_i>C_j$ the result holds immediately by Proposition \ref{prophighdelta}. Otherwise, let $\tilde{l}\geq2$ be the first instance such that $C_i^{\tilde{l}}>C_j^{\tilde{l}}$, hence also the first instance in which $F_i^{\tilde{l}}<F_j^{\tilde{l}}$. Hence:
\begin{align*}
\lim\limits_{\delta \to 1} \frac{DC_i(\delta)-DC_j(\delta)}{(1-\delta)^{\tilde{l}}}&=\lim\limits_{\delta \to 1}\frac{\left(\sum\limits_{l=1}^{n-1}\delta^lD_i^l\right)-\left(\sum\limits_{l=1}^{n-1}\delta^lD_j^l\right)}{(1-\delta)^{\tilde{l}}}\overset{\frac{0}{0}}{=}\\
&\overset{\frac{0}{0}}{=}\lim\limits_{\delta \to 1}\frac{\left(\sum\limits_{l=1}^{n-1}l\delta^{l-1}D_i^l\right)-\left(\sum\limits_{l=1}^{n-1}l\delta^{l-1}D_j^l\right)}{-\tilde{l}(1-\delta)^{\tilde{l}-1}}=\tag{\stepcounter{equation}\theequation}\\
&\overset{\frac{0}{0}}{=}\lim\limits_{\delta \to 1}\frac{\left(\sum\limits_{l=2}^{n-1}l(l-1)\delta^{l-2}D_i^l\right)-\left(\sum\limits_{l=2}^{n-1}l(l-1)\delta^{l-2}D_j^l\right)}{                                                                                                                                                                                                                                                                 \tilde{l}(\tilde{l}-1) (1-\delta)^{\tilde{l}-2}}=\\
&\overset{\frac{0}{0}}{=}\dots\overset{\frac{0}{0}}{=}\lim\limits_{\delta \to 1}\frac{\left(\sum\limits_{l=\tilde{l}}^{n-1}\frac{l!}{(l-\tilde{l})!}\delta^{l-\tilde{l}}D_i^l\right)-\left(\sum\limits_{l=\tilde{l}}^{n-1}\frac{l!}{(l-\tilde{l})!}\delta^{l-\tilde{l}}D_j^l\right)}{\tilde{l}!}=\\
&=\lim\limits_{\delta \to 1}\left(\sum\limits_{l=\tilde{l}}^{n-1}\frac{l!}{\tilde{l}!(l-\tilde{l})!}\delta^{l-\tilde{l}}D_i^l\right)-\left(\sum\limits_{l=\tilde{l}}^{n-1}\frac{l!}{\tilde{l}!(l-\tilde{l})!}\delta^{l-\tilde{l}}D_j^l\right)=-(F_i^{\tilde{l}}-F_j^{\tilde{l}})>0
\end{align*}
\noindent and by continuity with respect to $\delta$, there is $\overline{\delta}$ such that for all $\delta \in (\overline{\delta},1)$ it holds that $\frac{DC_i(\delta)-DC_j(\delta)}{(1-\delta)^{\tilde{l}}}>0$ or equivalently $DC_i(\delta)>DC_j(\delta)$.
\end{proof}

\begin{proof}[\textbf{\textup{Proof of Proposition~\ref{propintermediatedeltadomdeg}}}]
$DC_i(\delta)=\delta D_i+\sum\limits_{l=2}^{n-1}\delta^lD_i^l$, which gives $DC_i(1)=n-1$ for all $i$. Therefore, the polynomial $DC_i(\delta)-DC_j(\delta)$ has a root for $\delta=1$ and another for $\delta=0$, for any pair $i,j \in N$. Having observed that, we obtain the following expression.
\begin{align*}
DC_i(\delta)-DC_j(\delta)&=\delta\left[A_1+A_2\delta+\dots+A_{n-1}\delta^{n-2}\right]=\\
&=\delta(1-\delta)\left[-\sum\limits_{l=2}^{n-1}A_l-\sum\limits_{l=3}^{n-1}A_l\delta-\dots-(A_{n-1}+A_{n-2})\delta^{n-4}-A_{n-1}\delta^{n-3}\right]
\end{align*}
where $A_1=D_i-D_j$ and $A_l=D_i^l-D_j^l$ for $l \in \{2,\dots,n-1\}$, with the crucial observation being that $\sum\limits_{l=1}^{n-1}A_l=0$. This last equation allows us to rewrite the above expression as follows: 
\begin{equation}
DC_i(\delta)-DC_j(\delta)=\delta(1-\delta)\left[A_1+\sum\limits_{l=1}^2A_l\delta+\dots+\sum\limits_{l=1}^{n-3}A_l\delta^{n-4}+\sum\limits_{l=1}^{n-2}A_l\delta^{n-3}\right]
\label{dcdifA}
\end{equation}
From the last expression is apparent that a sufficient condition, though not necessary, to satisfy $DC_i(\delta)>DC_j(\delta)$ is that $\sum\limits_{l=1}^kA_l\geq0$ for all $k \in \{1,\dots,n-1\}$, with at least one inequality being strict. This in turn is equivalent to  $\sum\limits_{l=1}^kD_i^l\geq \sum\limits_{l=1}^kD_j^l$ for all $k$, which by definition means that $\mathcal{D}_i>_{UD}\mathcal{D}_j$.
\end{proof}

\begin{proof}[\textbf{\textup{Proof of Proposition~\ref{propintermediatedeltadomfar}}}]
Let $\epsilon=1-\delta$ and restate decay centrality as follows: $DC_i(\epsilon)=(1-\epsilon) D_i+\sum\limits_{l=2}^{n-1}(1-\epsilon)^lD_i^l$ which gives $DC_i(\epsilon=0)=n-1$ for all $i$. Therefore, the polynomial $DC_i(\epsilon)-DC_j(\epsilon)$ has a root for $\epsilon=1$ (equivalent to $\delta=0$) and another for $\epsilon=0$ (equivalent to $\delta=1$), for any pair $i,j \in N$. In addition to this, we consider the binomial identity: $(1-\epsilon)^l=\sum\limits_{k=0}^l(-1)^k \binom lk \epsilon^k$, which allows us to rewrite $DC_i(\epsilon)-DC_j(\epsilon)$ as follows:
\begin{align*}
DC_i(\epsilon)-DC_j(\epsilon)&=A_1\sum\limits_{k=0}^1(-1)^k\binom{1}{k} \epsilon^k+A_2\sum\limits_{k=0}^2(-1)^k\binom{2}{k} \epsilon^k+\dots+A_1\sum\limits_{k=0}^{n-1}(-1)^k\binom{n-1}{k} \epsilon^k=\\
&=\underbrace{\sum\limits_{l=1}^{n-1}A_l}_{=0}+\sum\limits_{l=1}^{n-1}A_l\binom{l}{1}(-\epsilon)+\sum\limits_{l=2}^{n-1}A_l\binom{l}{2}(-\epsilon)^2+\dots+\sum\limits_{l=n-1}^{n-1}A_l\binom{l}{n-1}(-\epsilon)^{n-1}=\\
&=-\epsilon\left[\sum\limits_{l=1}^{n-1}A_l\binom{l}{1}+\sum\limits_{l=2}^{n-1}A_l\binom{l}{2}(-\epsilon)+\dots+\sum\limits_{l=n-1}^{n-1}A_l\binom{l}{n-1}(-\epsilon)^{n-2}\right]=\\
&=-\epsilon\left[B_1+B_2\epsilon+\dots+B_{n-1}\epsilon^{n-2}\right]
\end{align*}
where $B_1=F_i-F_j$ and $B_l=F_i^l-F_j^l$ for $l \in \{2,\dots,n-1\}$. Having observed that and knowing that $\epsilon=1$ is also a root of the polynomial, we obtain the following expression via Euclidean division:
$$DC_i(\epsilon)-DC_j(\epsilon)=-\epsilon(1-\epsilon)\left[-\sum\limits_{l=2}^{n-1}B_l-\sum\limits_{l=3}^{n-1}B_l\epsilon-\dots-(B_{n-1}+B_{n-2})\epsilon^{n-4}-B_{n-1}\epsilon^{n-3}\right]$$
with the crucial observation being that $\sum\limits_{l=1}^{n-1}B_l=0$, which is the remainder of the division of the polynomial by $1-\epsilon$, which we know that it should be equal to zero. Hence, similarly to Proposition \ref{propintermediatedeltadomdeg}, we obtain the following expression: 
\begin{equation}
DC_i(\epsilon)-DC_j(\epsilon)=-\epsilon(1-\epsilon)\left[B_1+\sum\limits_{l=1}^2B_l\epsilon+\dots+\sum\limits_{l=1}^{n-3}B_l\delta^{n-4}+\sum\limits_{l=1}^{n-2}B_l\delta^{n-3}\right]
\label{dcdifB}
\end{equation}
It is apparent from the last expression that a sufficient condition, though not necessary, to satisfy $DC_i(\epsilon)>DC_j(\epsilon)$ for all $\epsilon \in (0,1)$, hence equivalently for all $\delta \in (0,1)$, is that $\sum\limits_{l=1}^kB_l\leq0$ for all $k \in \{1,\dots,n-1\}$, with at least one inequality being strict. This in turn is equivalent to  $\sum\limits_{l=1}^kF_i^l\leq \sum\limits_{l=1}^kF_j^l$ for all $k$, which by definition means that $\mathcal{F}_j>_{UD}\mathcal{F}_i$.
\end{proof}

\begin{proof}[\textbf{\textup{Proof of Proposition~\ref{propdeltalowerhalf}}}]
1. $DC_i(\delta)\geq DC_j(\delta) \Rightarrow D_i\delta\geq  D_j\delta+[(n-1)-D_j]\delta^2 \Rightarrow \delta\leq\frac{A_1}{n-1-D_j}$. The last inequality holds for all $\delta \in (0,1/2]$ if $2A_1\geq (n-1)-D_j$.

2.$DC_i(\delta)\geq DC_j(\delta) \Rightarrow D_i\delta+D_i^2\delta^2\geq  D_j\delta+D_j^2\delta^2+[(n-1)-D_j-D_j^2]\delta^3 \Rightarrow \delta\leq\frac{A_2+\sqrt{A_2^2+4A_1[(n-1)-D_j-D_j^2]}}{2[(n-1)-D_j-D_j^2]}$. The last inequality holds for all $\delta \in (0,1/2]$ if $A_1+2A_2\geq (n-1)-D_j-D_j^2$.

3. The result can also be implied by Rouch\'e's Theorem, but it is presented here with an independent proof, which follows a similar process to the one used for obtaining Cauchy's Bound of Polynomial Roots. More specifically, observe that by Proposition \ref{proplowdelta} together with $A_1>0$ it must hold true that $DC_i(\delta)-DC_j(\delta)=\delta\left(A_1+A_2\delta+\dots+A_{n-1}\delta^{n-2}\right)>0$ for $\delta$ close to zero and obviously has a root for $\delta=0$. We also know that it has another root for $\delta=1$. Then it is sufficient to show that $DC_i(\delta)-DC_j(\delta)$ has no other root for $\delta \in (0,1/2)$ as long as $A_1\geq |A_l|$ for all $l \in \{2,\dots,n-1\}$. To prove this, it is sufficient to focus on the term $\left(A_1+A_2\delta+\dots+A_{n-1}\delta^{n-2}\right)$.
\begin{align*}
\left|A_1+A_2\delta+\dots+A_{n-1}\delta^{n-2}\right| &\geq |A_1|-\left(|A_2|\delta+\dots+|A_{n-1}|\delta^{n-2}\right)=\\
&=|A_1|\left(1-\frac{|A_2|}{|A_1|}\delta-\dots-\frac{|A_{n-1}|}{|A_1|}\delta^{n-2}\right)\geq\\
&\geq |A_1|\left(1-\delta-\dots-\delta^{n-2}\right)=|A_1|\left(1-\sum\limits_{l=1}^{n-2}\delta^{l}\right)=\\
&=|A_1|\left(1-\frac{\delta-\delta^{n-1}}{1-\delta}\right)=|A_1|\frac{1-2\delta+\delta^{n-1}}{1-\delta}
\end{align*}

Therefore, the polynomial has the same number of roots in $(0,1)$ as $P(\delta)=1-2\delta+\delta^{n-1}$, which by Descartes' Rule of Signs has either zero or two positive roots. One of them is obsviously for $\delta=1$ and and the other is for some $\hat{\delta}<1$, because the polynomial has a unique minimum for some $\delta<1$ and is positive for $\delta=0$. Moreover, note that $P(1/2)>0$ for any $n$, and in fact $P(1/2)\to 0$ as $n \to \infty$. Hence, the expression is always positive in $(0,1/2]$, which turn means that $A_1+A_2\delta+\dots+A_{n-1}\delta^{n-2}$ has no root for $\delta \in (0,1/2]$ and in fact it is always positive in that region, by continuity.

4. The proof is identical to that of condition 3, if one considers the alternative expression of difference between decay centralities, $DC_i(\delta)-DC_j(\delta)$, that is provided by Equation (\ref{dcdifA}).
\end{proof}

\begin{proof}[\textbf{\textup{Proof of Proposition~\ref{propdeltaupperhalf}}}]
1. Consider again the reformulated expression with $\epsilon=1-\delta$, for which we  know that $DC_i(\epsilon)-DC_j(\epsilon)=-\epsilon\left[B_1+B_2\epsilon+\dots+B_{n-1}\epsilon^{n-2}\right]$, for which it is sufficient to show that it has no root for $\epsilon \in (0,1/2]$ as long as $|B_1|\geq|B_l|$ for all $l \in \{2,\dots,n-1\}$. This is proven identically to condition 3 of Proposition \ref{propdeltalowerhalf}. The obtained result, together with the fact that $B_1<0$ ensures that $DC_i(\epsilon)>DC_j(\epsilon)$ for all $\epsilon \in (0,1/2]$, which is identical to saying that $DC(\delta)>DC_j(\delta)$ for all $\delta \in [1/2,1)$.

2. The proof is identical to that of condition 1, if one consider the alternative expression of difference between decay centralities, $DC_i(\epsilon)-DC_j(\epsilon)$, that is provided by Equation (\ref{dcdifB}).
\end{proof}

\end{appendix}


\addcontentsline{toc}{section}{References}
\begin{thebibliography}{}

\bibitem[\protect\citeauthoryear{Ballester et.al}{2006}]{Ballesteretal2006} \textsc{Ballester, C., Calv\'o--Armengol, A. and Zenou, Y.} (2006). Who’s Who in Networks. Wanted: The Key Player. {\em Econometrica} 74, 1403--1417

\bibitem[\protect\citeauthoryear{Banerjee et.al}{2013}]{Banerjeeetal2013} \textsc{Banerjee, A., Chandrasekhar, A.G., Duflo, E. and Jackson, M.} (2013). The Diffusion of Microfinance. {\em Science} 341 (6144).

\bibitem[\protect\citeauthoryear{Bloch et.al}{2016}]{Blochetal2016} \textsc{Bloch, F., Jackson, M.O. and Tebaldi, P.} (2016).Centrality Measures in Networks. {\em SSRN Working Paper}.

\bibitem[\protect\citeauthoryear{Bonacich}{1987}]{Bonacich1987} \textsc{Bonacich, P.} (1987). Power and Centrality: A Family of Measures. {\em American Journal of Sociology} 92, 1170--1182.

\bibitem[\protect\citeauthoryear{Brandes and Erlebach}{2005}]{BrandesErlebach2005} \textsc{Brandes, U. and Erlebach, T.} (2005). \textit{Network Analysis: Methodological Foundations}. Springer.

\bibitem[\protect\citeauthoryear{Chatterjee and Dutta}{2015}]{ChatterjeeDutta2015} \textsc{Chatterjee, K. and Dutta, B.} (2015). Credibility and Strategic Learning in Networks. {\em International Economic Review}, forthcoming.

\bibitem[\protect\citeauthoryear{Dequiedt and Zenou}{2014}]{DequiedtZenou2014} \textsc{Dequiedt, V. and Zenou, Y.} (2014). Local and Consistent Centrality Measures in Networks. {\em CEPR Discussion Paper No. DP10031}.

\bibitem[\protect\citeauthoryear{Dijkstra}{1959}]{Dijkstra1959} \textsc{Dijkstra, E.W.} (1959). A Note on Two Problems in Connexion with Graphs. {\em Numerische Mathematik} 1, 269--27.

\bibitem[\protect\citeauthoryear{Erd{\H o}s and R\'enyi}{1959}]{ErdosRenyi1959} \textsc{Erd{\H o}s, P. and R\'enyi, A.} (1959). On Random Graphs I. {\em Publ. Math. Debrecen} 6, 290--291.

\bibitem[\protect\citeauthoryear{Faust}{1997}]{Faust1997} \textsc{Faust, K.} (1997). Centrality in Affiliation Networks. {\em Social Networks} 19, 157--191.

\bibitem[\protect\citeauthoryear{Galeotti and Goyal}{2009}]{GaleottiGoyal2009} \textsc{Galeotti, A. \& Goyal, S.} (2009). Influencing the Influencers: a Theory of Strategic Diffusion. {\em RAND Journal of Economics} 40(3), 509--532.


\bibitem[\protect\citeauthoryear{Gofman}{2015}]{Gofman2015} \textsc{Gofman, M.} (2015). Efficiency and Stability of a Financial Architecture with Too--Interconnected--To--Fail Institutions. {\em Journal of Financial Economics}, forthcoming.

\bibitem[\protect\citeauthoryear{Katz}{1953}]{Katz1953} \textsc{Katz, L.} (1953). A New Status Index Derived from Sociometric Analysis. {\em Psychometrica}, 18, 39--43.

\bibitem[\protect\citeauthoryear{K{\"o}nig et.al}{2014}]{Konigetal2014} \textsc{K{\"o}nig, M., Tessone, C.J. and Zenou,Y.} (2014). Nestedness in Networks: A Theoretical Model and Some Applications. {\em Theoretical Economics} 9, 695-752.

\bibitem[\protect\citeauthoryear{Jackson}{2008}]{Jackson2008} \textsc{Jackson, M.O.} (2008). \textit{Social and Economic Networks}. Princeton University Press.

\bibitem[\protect\citeauthoryear{Jackson}{2016}]{Jackson2016} \textsc{Jackson, M. O.} (2016). The friendship Paradox and Systematic Biases in Perceptions and Socal Norms. {\em SSRN Working Paper}.

\bibitem[\protect\citeauthoryear{Jackson and Wolinsky}{1996}]{JacksonWolinsky1996} \textsc{Jackson, M. O. and Wolinsky, A.} (1996). A Strategic Model of Social and Economic Networks. {\em Journal of Economic Theory}, 71, 44--74.

\bibitem[\protect\citeauthoryear{Liu et.al}{2015}]{Liuetal2015} \textsc{Liu, X., Patacchini, E., Zenou, Y. and Lung--Fei, L.} (2015). Criminal Networks: Who is the Key Player? working paper.

\bibitem[\protect\citeauthoryear{Palacios--Huerta and Volij}{2004}]{PalaciosHuertaVolij2004} \textsc{Palacios--Huerta, I. and Volij, O.} (2004). The Measurement of Intellectual Influence. {\em Econometrica}, 72, 963--977.

\bibitem[\protect\citeauthoryear{Pastor--Satorras and Vespignani}{2002}]{PastorSatorrasVespignani2002} \textsc{Pastor--Satorras, R. and Vespignani, A.} (2002). Immunization of Complex Networks . {\em Physical Review E}, 65: 036104.

\bibitem[\protect\citeauthoryear{Rothenberg et.al}{1995}]{Rothenbergetal1995} \textsc{Rothenberg, R., Potterat, J.J., Woodhouse, D.E., Darrow, W.W., Muth, S.Q. and Klovdahl A.S.} (1995). Choosing a centrality measure: Epidemiologic correlates in the Colorado Springs study of social networks. {\em Social Networks}, 17, 273--297.

\bibitem[\protect\citeauthoryear{Tsakas}{2016}]{Tsakas2016} \textsc{Tsakas, N.} (2016). Optimal Influence under Observational Learning. {\em SSRN Working Paper}.

\bibitem[\protect\citeauthoryear{Valente et.al}{2008}]{Valenteetal2008} \textsc{Valente, T. W., Corognes, K., Lako, C. and Costenbader, E.} (2008). How Correlated are Network Centrality Measures. {\em Connect (Tor)}, 28, 16--26.


\end{thebibliography}
\end{document}